\documentclass[pdflatex,sn-nature]{sn-jnl}
\usepackage[T1]{fontenc}
\usepackage{xcolor}
\usepackage{lmodern,amsmath,amssymb,graphicx,booktabs,array,longtable,microtype,placeins,xurl, lineno}
\DeclareGraphicsExtensions{.pdf,.png}
\setcitestyle{super,comma}

\newcommand{\paperfig}[3]{\begin{figure}[!htbp]\centering\includegraphics[width=\textwidth,height=.59\textheight,keepaspectratio]{figures/#1}\caption{#2}\label{#3}\end{figure}\FloatBarrier}

\begin{document}
\title{{\huge How fault interactions change large-earthquake probability during the seismic cycle}}
\author*[1,2]{\fnm{Davide} \sur{Zaccagnino}}\email{zaccagnino@sustech.edu.cn}
\author*[3]{\fnm{Laura} \sur{Gulia}}
\affil[1]{Institute of Risk Analysis, Prediction and Management (Risks-X), Academy for Advanced Interdisciplinary Studies, Southern University of Science and Technology, Shenzhen, China}
\affil[2]{National Institute of Geophysics and Volcanology (INGV), Rome, Italy}
\affil[3]{Department of Earth Sciences, University of Pisa, Pisa, Italy}

\abstract{Most current seismic forecasts estimate variations in large-earthquake probability from the seismic rate alone. Unless foreshocks raise that rate, a large rupture can appear as improbable on the eve of a mainshock as during background seismicity. This assumes that the distribution of rupture sizes does not change as stress accumulates. However, as more fault patches get closer to failure, the interaction regions susceptible to perturbations become larger. We derive how this increase in susceptibility allows more ruptures to overcome barriers, lowering the $b$-value of the Gutenberg-Richter law and raising the corner magnitude, \(M_x\). Physical fault simulations reproduce these trends. Earthquake-triggering simulations show that a rise in $M_x$ can produce an apparent decrease in $b$, but a joint monitoring can resolve their contributions. At Ridgecrest and L’Aquila, $M_x$ slightly increases before the mainshocks and clearly decreases during aftershocks. For earthquake magnitudes above the initial corner, a shift in $M_x$ dominates the change in exceedance probability. Therefore, monitoring $M_x$ with $b$ jointly is crucial to give forecasts a physical measure of the evolving capacity of faults to generate large earthquakes.}

\maketitle

\section*{Why are large earthquake probabilities always small?}
Reliable short- and mid-term earthquake forecasting requires understanding how stress accumulation changes the sizes of ruptures that faults can generate over time. 

Most magnitude models, however, assume that this size distribution is fixed.

They rely on the Gutenberg-Richter (GR) law\cite{gutenberg1944,ishimoto1939}: for a chosen region and time interval, the cumulative number of earthquakes with magnitude at least \(M\), denoted by \(N(\ge M)\), follows
\begin{equation}
	\log_{10}N(\ge M)=a-bM.
	\label{eq:GR_intro}
\end{equation}
Here, \(a\) describes the regional productivity and \(b\) determines the relative abundance of small and large events. Above a completeness threshold \(M_c\), the probability that an event exceeds \(M\) is given by \(S_{\rm GR}(M)=10^{-b(M-M_c)}\). For \(b=1\), each additional magnitude unit reduces this probability tenfold. 

If the magnitude distribution was time-invariant, the event rate would be the only quantity that controls the probability of a large earthquake. In this description, stress accumulation progressively drives fault systems closer to instability but, paradoxically, the rate-rescaled probability of large earthquakes is constant.

The widespread occurrence of frequency-magnitude power-law scaling has led to theories predicting a universal exponent\cite{kagan1999,kagan2010}. Specifically, critical-branching and mean-field models provided a first comprehensive description of the seismic behavior of fault systems in the long term and over wide regions, implying constant size-distribution exponents\cite{fisher1997,sethna2001}. 

Observations, however, show systematic variations of the \(b\)-value in space and time. 
Spatial patterns mostly depend on faulting style and stress\cite{wiemer1997,schorlemmer2005,scholz2015}; this interpretation is supported by laboratory experiments revealing changes during loading, repeated failure and other features of faulting\cite{scholz1968,amitrano2003,goebel2013, zaccagnino2022scaling}. Earthquake sequences also show changes in \(b\) before and after major ruptures\cite{nanjo2012,gulia2010,gulia2020}. 
In the field of seismic hazard, Gulia and Wiemer used changes in the \(b\)-value after a strong earthquake to assess whether a larger shock might follow\cite{gulia2019}. 

This evidence and applications require a physical explanation of how the magnitude distribution evolves during the seismic cycle, and solicits reconciliation with the universal picture of point-process models.

Specifically, the observational evidence of a decreasing slope may have different physical origins: does loading change the ratio of rupture sizes within the power-law range, extend that range to larger possible magnitudes, or do both?

Indeed, in disordered systems, power-law event-size distributions arise from cascades of interacting failures spanning a range of spatial scales\cite{fisher1997}. A cascade can involve only those patches whose failures are made possible through dynamic stress transfer\cite{mignan2014debate}, so it remains confined to an interaction region whose spatial extension is measured by its correlation length, $\xi$. This value depends on both stress transfer and the susceptibility of the receiving patches: elastic perturbations can propagate far from their source, but their triggering potential depends on the surrounding fault state\cite{benzion2008,kato2021} and evolve during the seismic cycle\cite{zaccagnino2021stress}. As stress approaches the failure threshold, more patches become susceptible to triggering\cite{brodsky2014triggering}, allowing cascades to involve progressively larger regions\cite{zaccagnino2024foreshocks, zaccagnino2026foreshocks}. Within this framework, the range of rupture sizes displaying approximate GR scaling extends towards larger magnitudes. Observational studies have investigated increasing spatial correlations before large earthquakes and tested their association with subsequent seismicity with mixed results\cite{zoller2001correlation,zoller2002critical}. Numerical fault models also show the development and disruption of seismic interactions during the seismic cycle\cite{benzion2003criticality}.

Therefore, a physical description of the seismic cycle must address how stress changes both the hierarchy of rupture sizes and the extent of cascading failures. 
The slope, represented by the $b$-value, grasp the hierarchy of fault activation within the scaling range. Changes in fault connectivity and slip scaling can modify it. 
At the same time, the characteristic extent of interacting failures leads to the spontaneous emergence of upper rupture-size scales, which determines an exponential cut-off in the event frequency-magnitude distribution, e.g., \cite{zaccagnino2025santorini, roche2026probability, rodriguez2026role}. 
The tapered Gutenberg-Richter (TGR) distribution represents this departure through the corner magnitude \(M_x\)\cite{kagan2002,bird2004}. The taper progressively reduces large-event probabilities relative to GR. At fixed \(b\), increasing \(M_x\) weakens the suppression and makes larger earthquakes more probable. Although \(b\) and \(M_x\) describe different properties, both affect a slope estimated by fitting ordinary power-laws to a tapered distribution. Physically, the expansion of the interaction regions raises $M_x$, making larger earthquakes more probable, and vice versa.

Tracking both the $b$-value and $M_x$ provides a means of assessing how stress accumulation and release modulate the range of rupture sizes that faults can produce, including changes in large-earthquake probability that seismic rate or $b$ alone would miss.

Here, we first derive how the spatial extent of interacting fault patches affects the probability of exceeding a given magnitude at a fixed seismicity rate. Physical simulations of heterogeneous faults test whether increasing stress accumulation produces the analytically predicted changes in \(b\) and \(M_x\). 
Then, we employ Epidemic-type aftershock sequence (ETAS) simulations, in which both quantities are known, to test whether their effects can be separated in simulated catalogs. 
Finally, we ask to which degree their evolution is detectable pseudo-prospectively in real catalogs to improve large earthquake forecasts. 
We show that an evolving corner magnitude adds predictive value supporting the theoretical prediction that the seismic potential of faults changes during the seismic cycle, with a minimum at the end of aftershock sequences and a maximum when mainshocks occur.

\section*{How fault stress modulates rupture sizes and earthquake probabilities}
The physical picture introduced above describes earthquake occurrence within a structurally heterogeneous fault systems controlling nucleation, propagation and arrest. 
An earthquake begins when a patch fails, redistributing stress to its surroundings. Whether the rupture stops or spreads depends on the strengths of the surrounding segments, their proximity to failure and the stress transferred to them. The resulting event releases a seismic moment determined by its rupture area and slip. 
The GR slope describes how rapidly earthquake numbers decrease as rupture moment increases. Therefore, its value depends on both the abundance of available rupture paths at different lengths and the moment released at each length. Let $L$ denote the rupture length and $t$ time. We describe the cumulative length distribution by $N(\ge L,t)\propto L^{-D(t)}$ and the moment-length relation by $M_0\propto L^{\rho(t)}$. Here, $D$ describes the hierarchy of rupture lengths\cite{hirata1989}, while $\rho$ accounts for rupture area and slip. Using the moment--magnitude relation\cite{hanks1979} gives
\begin{equation}
	b(t)=\frac{3D(t)}{2\rho(t)}.
	\label{eq:intrinsicb}
\end{equation}
A fault system that produces proportionally more long ruptures has a smaller $D$. A system in which moment increases more rapidly with rupture length has a larger $\rho$. Both changes lower $b$. For self-similar ruptures with constant stress drop, $\rho=3$ and $b=D/2$\cite{aki1981}. The key point is that the slope reflects how ruptures occupy the fault system and how slip develops as they spread (see the Supplementary Section~\ref{sec:stress} for details). This result is well known in avalanche theories which show how dimensionality and roughness enter event-size scaling beyond mean-field limits\cite{ledoussal2009,rosso2009}.

The same propagation process determines the spatial range over which this rupture hierarchy develops. 
When the surrounding patches are far from failure, a stress perturbation can trigger only a short cascade. As loading brings more patches closer to their strength thresholds, the same perturbations can sustain failure over larger regions. This increasing spatial extent can be measured by a correlation length, $\xi$ which grows as the effective fault stress, $\Sigma(t)$, approaches the threshold for collective instability, $\Sigma_c$. 
In ideal critical systems, the critical limit corresponds to a diverging $\xi$; however, natural fault systems have finite sizes and structural barriers constrain its growth in the crust, which introduces a taper in the frequency-size distribution of ruptures. 

Nevertheless, an expanding interaction region allows a cascade to involve a larger rupture area and release a greater seismic moment. The magnitude scale associated with the suppression of large events must then shift accordingly. We represent this scale by the corner magnitude\cite{kagan2002}, $M_x$. Within the critical scaling regime, its dependence on stress follows
\begin{equation}
	M_x(t)=M_x(0)-A_x\log_{10}
	\left[\frac{\Sigma_c-\Sigma(t)}
	{\Sigma_c-\Sigma_0}\right].
	\label{eq:stresscorner}
\end{equation}
Here, $\Sigma_0=\Sigma(0)$ is a reference stress below $\Sigma_c$, and $A_x$ is a positive quantity describing the sensitivity of the corner magnitude to the stress gap. Each tenfold reduction in this gap raises $M_x$ by $A_x$ magnitude units. Supplementary Section~\ref{sec:regimes} provides the derivation and related assumptions. 

This relation describes an increasing potential for large rupture as collective failure stress approaches. It is interesting to notice that even under steady loading, the increase in $M_x$ accelerates as the stress gap narrows. At the same time, finite fault dimensions limit this progression. Conversely, following an earthquake, stress release can reduce susceptibility within the ruptured region, while transferred stress can promote cascading failure nearby. Therefore, relaxation and reloading produce both decreasing and increasing $M_x$ during the seismic cycle while modulating the $b$-value: the two processes are part of the same physical dynamics and cannot be separated. Supplementary Section~\ref{sec:regimes} reports the analytical responses for background, foreshock and aftershock activity under suitable hypotheses.

These results show that stress can change both the slope of the GR law and the magnitude range over which it applies. 
Let $M_c$ denote the completeness magnitude threshold, $M_0(M)$ the seismic moment corresponding to moment magnitude $M$, and $S$ the probability that an event above $M_c$ exceeds $M$. 
Then
\begin{equation}
	S(M\mid b,M_x)=10^{-b(M-M_c)}
	\exp\!\left[
	-\frac{M_0(M)-M_0(M_c)}{M_0(M_x)}
	\right]. 
	\label{eq:mainTGR}
\end{equation}
This expression applies for $M\ge M_c$. The first factor describes the relative abundance of rupture sizes within the power-law GR range. The second represents the progressive suppression of large ruptures associated with the finite size of stress interaction regions. Increasing $M_x$ weakens this suppression, raising large-earthquake probabilities even when $b$ stays constant. The effect can be particularly strong for target magnitudes above the initial corner.

Potentially - it is worthwhile to notice - the taper also influences a slope estimated using ordinary GR; indeed, strong suppression reduces the mean magnitude and steepens the estimated slope (see Supplementary Section~\ref{sec:bias}). 
As $M_x$ increases, that estimate can decrease even without a change in the rupture hierarchy. Therefore, interpreting an observed $b$ decrease requires assessing both the slope and the taper. When the target magnitude is above the initial corner, increasing $M_x$ can raise large-earthquake probabilities by orders of magnitude and contribute more than a decrease in $b$. Supplementary Section~\ref{sec:probability} derives and compares these contributions.

Together, these results describe two physical effects acting on the magnitude distribution: while fracture-length statistics and slip scaling determine the $b$-value, the spatial extent of stress-correlated fault networks affects the $M_x$. Stress changes both the organization of ruptures and the spatial extent over which failure can spread modulating large-earthquake probability, and jointly making them necessary for interpreting the actual fault state. 

\section*{Interaction regions evolve during the seismic cycle in natural settings}
We investigate whether natural seismicity shows the variations in $b$ and $M_x$ expected from the physical framework developed above. 

The region of Ridgecrest, California, offers a high-quality sequence of foreshocks, a major rupture and abundant aftershocks in a complex fault system\cite{ross2019,atterholt2025}. Its $M_w 6.4$ foreshock preceded the $M_w 7.1$ mainshock by about 34 hours. The dense Southern California Earthquake Catalogue allows us to follow changes in the magnitude distribution before and after these events, with a finer temporal resolution during periods of intense activity. 
We also analyse L'Aquila, where an extended foreshock sequence preceded the 2009 $M_w 6.3$  mainshock in a contrasting fault setting. The corresponding results are presented in Supplementary Section~\ref{sec:laquila}.

For monitoring the temporal changes of the frequency-size distribution of magnitudes, we introduce the evolving tapered Gutenberg-Richter (ETGR) model, where we estimate $b$ and $M_x$ before each earthquake using up to 500 earlier events. The forecast includes the estimation of the uncertainties to assess whether these observations support tapering, allowing the prediction to approach ordinary GR when the upper magnitude scale is poorly constrained (Methods). We then evaluate the probability assigned to the next magnitude. Repeating this calculation through the catalog allows us to test our predictions against subsequent observations, following the likelihood-based approach\cite{schorlemmer2007testing}.

As a first step, we assess the predictive performance employing the epidemic-type aftershock sequence (ETAS) framework. ETAS provides a widely used benchmark for short-term earthquake forecasting: it describes background activity and the cascades of earthquakes triggered by earlier events\cite{ogata1988,werner2011}. We calibrate its occurrence-rate using seismicity, then compare alternative magnitude distributions using the same dataset. This comparison tests whether temporal changes of the $b$-value and corner magnitude $M_x$ provide predictive information in addition to the changing activity rate that ETAS estimates. Figure~\ref{fig:forecast} assess their performance in assigning probabilities to the magnitudes of subsequent earthquakes.
It compares the evolving magnitude distribution (ETGR) at Ridgecrest with a constant TGR description of the full catalogue. To assess its predictive performance, we estimate each updating model using only earlier earthquakes and evaluate the probability it assigns to the next observed magnitude (see Methods). Panel D shows the cumulated differences between these predictive logarithmic scores. Most of the gain develops during aftershock recovery, when abundant seismicity provides greater temporal resolution. We also implement an ETGR which allows unbounded $M_x$ estimates (hence reducing TGR to GR) which achieves comparable or slightly better scores. This result shows that allowing the magnitude distribution to evolve both in its scaling and cut-off components supplies a clear predictive improvement in this sequence.

\paperfig{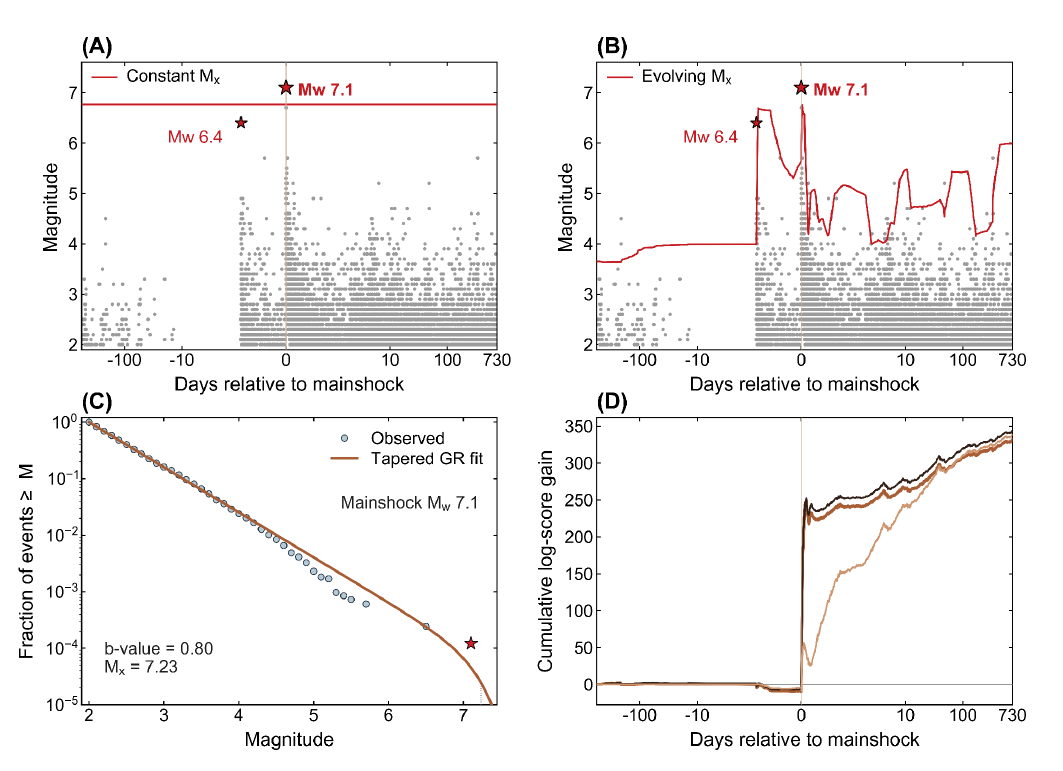}{\textbf{ETAS investigation of the evolving earthquake-size distribution at Ridgecrest.} Southern California Earthquake Data Center catalog, July 2014-July 2021, for magnitudes $\ge2.0$ within 60\,km of the mainshock and depths of 0-30\,km. ETAS forecast evaluation starts from January 2018. \textbf{A)} Catalog magnitudes, with the constant corner from the pre-mainshock catalogue TGR fit in red. \textbf{B)} Corner magnitude estimated from up to 500 earlier earthquakes, shown as a moving average of the 200 previous estimates. \textbf{C)} Fractions of events exceeding each magnitude, with observations in blue and the TGR fit in copper. The dotted line marks $M_x = 7.23$. Red stars identify the $M_w 6.4$ foreshock and $M_w 7.1$ mainshock. \textbf{D)} Cumulative predictive log-score differences between the calibrated ETAS models using different best-fitting frequency-size distributions: evolving TGR (ETGR) versus constant TGR (copper), evolving TGR with possible unbounded $M_x$ (mixture TGR) versus constant TGR (dark brown), and ETGR versus a reference TGR using all previous events (light copper). Positive values favor the evolving models, with most gains during aftershock recovery. Time is relative to the July 6, 2019 mainshock, with logarithmic horizontal axis.}{fig:forecast}

The comparison of predictive skills in Figure~\ref{fig:forecast} shows that updating the magnitude distribution improves the estimates of the probabilities of future earthquakes with a clear information gain in aftershocks sequences. 

Then, the aim is to reconstruct the evolution of the magnitude distribution in real catalogs by fitting data over time. We study how its slope and taper evolve through seismic activity and how these changes are related to the fault system (Fig.~\ref{fig:cycle}) using the same regions and directly analysing the catalog. To do this, we select seven key periods to better highlight differences in the various phases of the seismic cycle: background activity, the final year before the sequence, foreshocks and four stages of aftershocks. Their magnitude distributions show how the range of observed earthquake sizes changes significantly between these stages. Estimates updated after each event provide a high-resolution temporal view of the evolution of $b$ and $M_x$. 
The foreshock interval and earliest aftershocks clearly show a broader upper tail than the background period, followed by changes in both slope and taper during recovery (Fig.~\ref{fig:cycle}B). These differences demonstrate why a magnitude distribution with fixed $b$-value and corner magnitude provides an incomplete description of the sequence.
The sequence combines lower $b$-values with higher corner estimates, followed by an increase of the slope and significant variations in $M_x$ (Fig.~\ref{fig:cycle}C,D). 

\paperfig{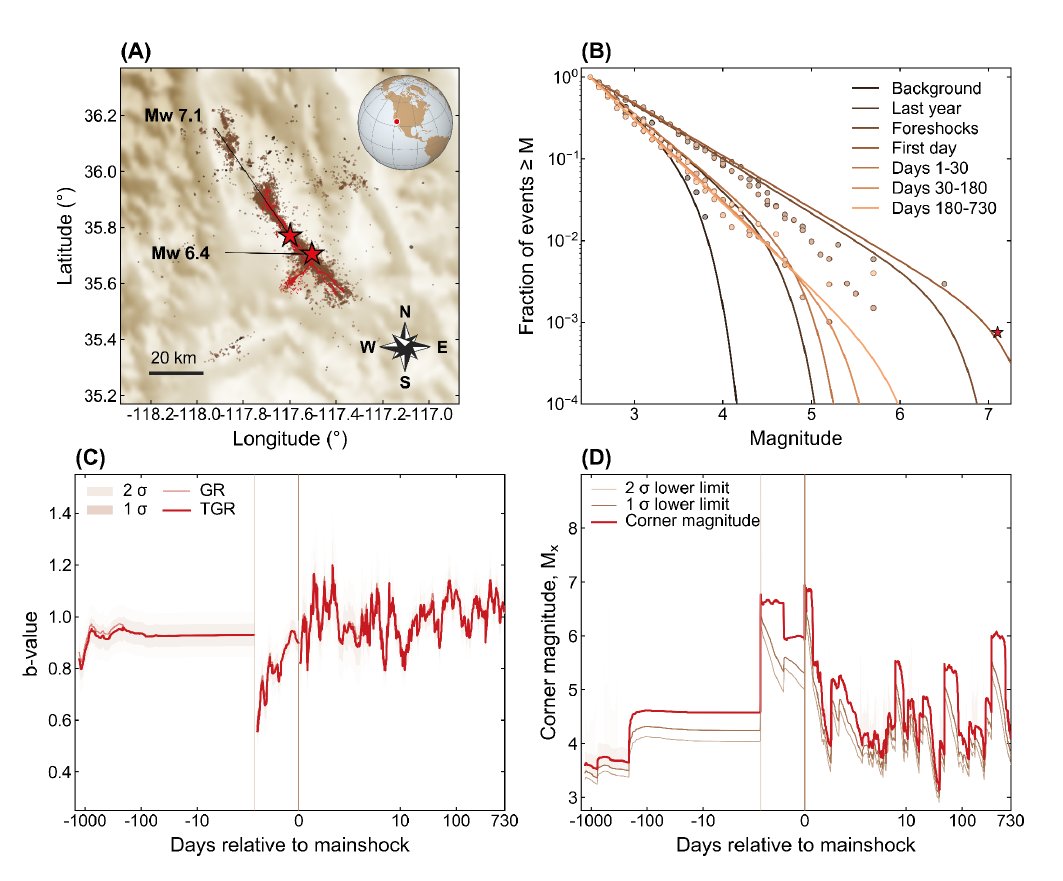}{\textbf{Magnitude distributions and their parameters at Ridgecrest.} Catalog, spatial selection and study interval follow Figure~\ref{fig:forecast}. Distributions use magnitudes above 2.5. \textbf{A)} Map of events in catalog, with mapped surface ruptures in red\cite{ponti2020}. Red stars mark the $M_w 6.4$ foreshock and $M_w 7.1$ mainshock. \textbf{B)} Complementary cumulative fractions of magnitude and corresponding TGR fits, progressing from dark to light copper and representing seven consecutive periods: July 2014-July 2018, the following year up to the $M_w 6.4$ foreshock, the interval between the two large earthquakes, and recovery during the first day, days 1-30, days 30-180 and days 180-730. The mainshock appears at $M_w 7.1$. \textbf{C)} GR and TGR $b$-values in pale and dark red. Darker and lighter shading indicates $1\sigma$ and $2\sigma$ uncertainties. Results are consistent with the output for the broader selected regions around the mainshock reported in \cite{gulia2020}, whereas the near-fault (in-box) analysis shows higher b-values. \textbf{D)} Corner magnitude is in red, with lower limits in copper and shading where both upper limits are rigid. Vertical lines mark the two largest earthquakes. Time is relative to the mainshock, logarithmic scaling.}{fig:cycle}

L'Aquila shows corresponding changes during its foreshock sequence and subsequent stabilization (Supplementary Figs.~\ref{fig:s1} and~\ref{fig:s2}). We follow both sequences for two years after their mainshocks to get a common observation period. The predictive tests and observational fits are used to provide complementary information: while the former show that recent magnitude statistics improve the probabilities assigned to future events, the latter reveal how the rupture hierarchy and upper magnitude scale evolve through the sequence. 
Within our physical framework, we use $b$ to describe the hierarchy of rupture sizes and $M_x$ as a statistical proxy of the interacting spatial regions. 
Our ETGR model provides a practical way to monitor both quantities through earthquake catalogs, with seismicity density controlling temporal resolution and uncertainty. This model quantifies and supports through two observables our physical hypothesis: stress accumulation and release change the range of rupture sizes that a fault system can generate.

\section*{Physical simulations distinguish hierarchical and extreme statistical modulations}
At Ridgecrest, the magnitude distribution flattened and extended towards larger events between the \(M_w 6.4\) foreshock and the \(M_w 7.1\) mainshock (Fig.~\ref{fig:cycle}). Similar or opposite patterns are observed in other phases of the seismic cycle at Ridgecrest and L'Aquila and have been widely documented in other settings\cite{gulia2019}. These observations need to be interpreted together, because a shift of the taper towards larger magnitudes can itself lower the slope estimated under GR (see Supplementary Section~\ref{sec:bias}). In general, the observed maximum magnitude and the GR slope are negatively correlated; therefore, it can be challenging to assess their separate contributions.

To quantify them, we use two different kinds of simulations. 

First, we generate synthetic ETAS catalogues, in which \(b\) and \(M_x\) can be varied separately without affecting the clustering characteristic of triggered seismicity\cite{helmstetter2003}. 
When only \(M_x\) increases, the ordinary GR estimate falls, but TGR recovers the rise in \(M_x\) together with a nearly constant intrinsic slope (Supplementary Fig.~\ref{fig:s5}). 
Conversely, when only \(b\) decreases, the model comparison identifies the slope change in most realisations. In simulations where the taper progressively moves beyond the sampled range, the evolving distribution also assigns higher cumulative probabilities to subsequent magnitudes than the initial calibration does, as measured by the event-by-event log score (Supplementary Section~\ref{sec:bias}). 

Second, we perform physics-based simulations of a rough fractal fault with heterogeneous strength and increasing differential stress. We record the ruptures that spontaneously occur when the local threshold is overtopped, with their timing and magnitude, and we then study how the \(b\)-value and the corner magnitude change. We increase the applied stress until a patch slips, then hold the load fixed throughout the ensuing cascade. Stress transfer acts between neighbouring patches, so the distance a rupture travels depends on how far successive failures carry it. Small events persist throughout loading, but a progressively larger fraction of cascades reach high magnitudes as failure approaches (Fig.~\ref{fig:fault}C,D).

\paperfig{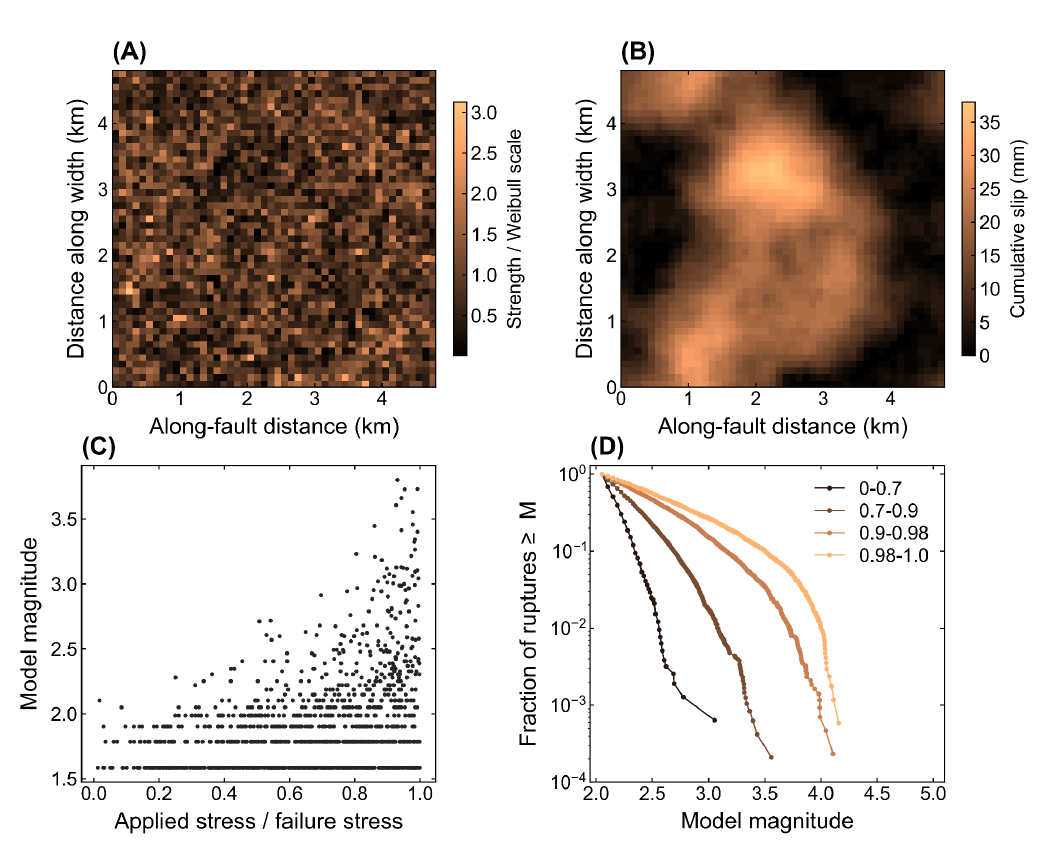}{\textbf{Physical simulation of seismicity on a structurally heterogeneous fractal fault.} \textbf{A)} Initial strength on a 4.8\,km fault divided into 100\,m patches. Lighter shades denote stronger patches, with strength expressed relative to the Weibull scale. \textbf{B)} Example of fault slip distribution in millimetres during a large seismic event. \textbf{C)} Simulated magnitudes as a function of applied stress divided by failure stress. \textbf{D)} Frequency--magnitude distributions at four different loading times, from early loading (dark brown) to near failure (light copper). Each curve is normalised by the number of ruptures in its interval, so their separation shows a change in the relative abundance of large events.}{fig:fault}

The fault shows both a lowering in the \(b\)-value and a rise of nearly two magnitude units in \(M_x\) between early loading and failure (Fig.~\ref{fig:stack}A,B). The drop of \(b\) persists also when the fit includes the taper, indicating a change in the rupture hierarchy within the scaling range. 
We perform upscaling analysis to investigate the effect of fault dimensions: faults of different widths have similar corners early in loading, but their corners separate as failure approaches, with the widest fault reaching the highest \(M_x\) (Fig.~\ref{fig:stack}C). 
Most importantly for the physical interpretation, the corner also rises when we calculate event size from the area of different patches that slipped (Fig.~\ref{fig:stack}D). Therefore, larger ruptures near failure involve a greater area of the fault, even after removing the contribution of repeated slip on the same patches.
For the strength distribution used here, a given stress increment is more likely to trigger a neighboring patch at higher background stress. Cascading transfers can then be enhanced and allow slip over a larger region, even though the range of each transfer is fixed\cite{benzion2008}. As the region ready to fail in a single cascade expands, the taper shifts towards larger magnitudes, increasing the probability that a small instability degenerate into a large earthquake.

\paperfig{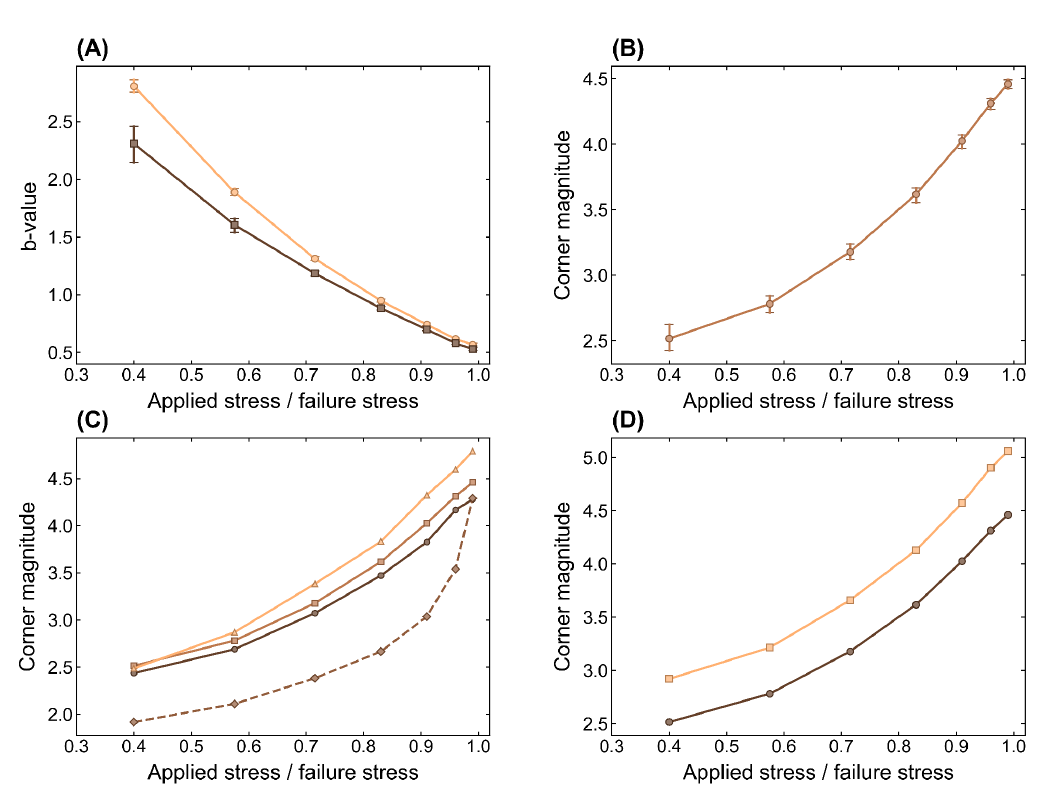}{
	\textbf{Evolution of the rupture hierarchy and corner magnitude during fault loading.}
	The horizontal axis gives applied stress as a fraction of the failure stress in each run.
	\textbf{A)} GR (light copper circles) and TGR (dark brown squares) \(b\)-values.
	\textbf{B)} Corresponding TGR corner magnitude. Error bars in \textbf{A)} and \textbf{B)} span the central 95\% of estimates obtained by resampling whole runs.
	\textbf{C)} Corner magnitudes for faults 10, 13 and 20\,km wide (circles, squares and triangles, respectively). Diamonds on the dashed curve show the 13\,km fault with spatially uncorrelated strengths.
	\textbf{D)} Corner magnitudes calculated from total slip (dark brown circles) and rupture area (light copper squares). Both increase towards failure.
}{fig:stack}

\section*{Monitoring rupture potential unveils varying large-earthquake probability}
Theory, observations and simulations show how stress accumulation flattens the hierarchy of ruptures and the expands the regions over which fault patches interact. 
This process progressively increases the largest magnitude, while changes in fracture statistics shifts the balance between small and large events.

Figure~\ref{fig:concept} summarizes this picture. 

\paperfig{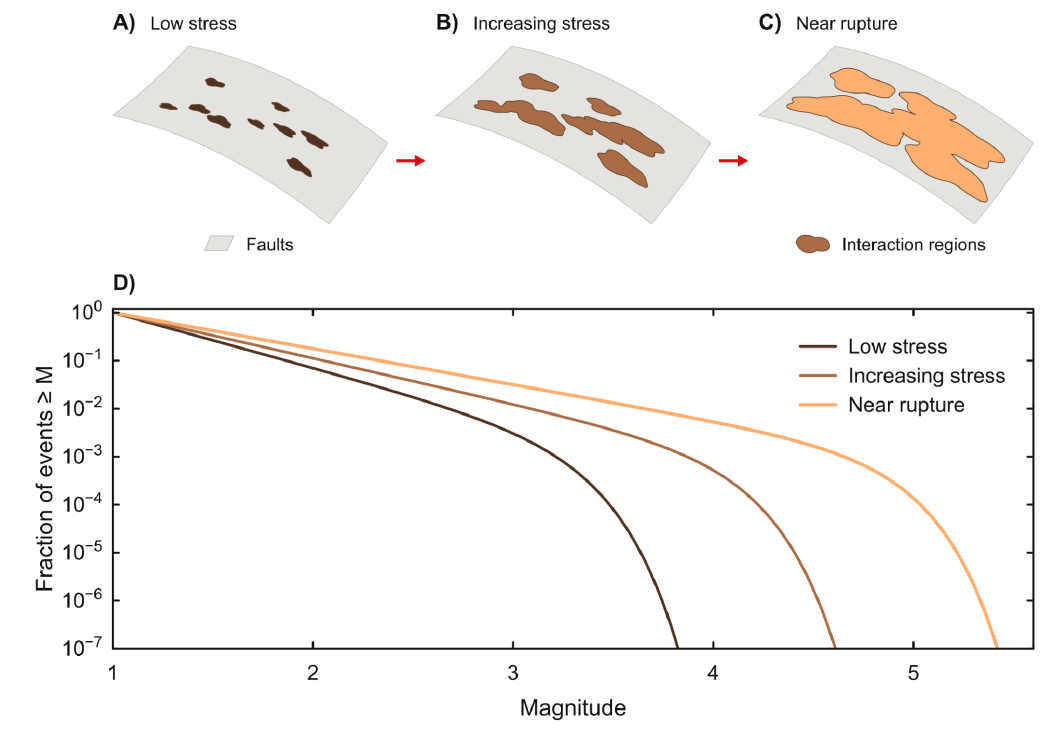}{\textbf{Physical interpretation of evolving rupture scales.} \textbf{A), B), C)} Schematic interaction regions on a curved fault under increasing stress. Small irregular regions grow and connect, allowing stress transfer to coordinate larger ruptures. Dark-to-light copper regions represent increasing stress, and simple red arrows indicate the progression towards failure. \textbf{D)} Illustrative TGR distributions show the corresponding extension of the upper tail. The corner controls the scale at which the distribution bends down, while the $b$-value controls its power-law slope. An increasing corner raises the relative frequency of large earthquakes; changes in rupture statistics reduce the $b$-value.}{fig:concept}

The two effects act on different parts of the magnitude distribution: the corner controls where the distribution bends down, while the \(b\)-value defines its slope within the scaling range. Therefore, their relative importance for hazard depends on the value of the target magnitude, and neither can substitute for the other. This is why they should be studied together. 

The principal hazard implication is that stress accumulation can change the range of rupture sizes with a probability that can be measured. As interaction regions expand and merge, the taper moves towards larger magnitudes and its suppression of large events weakens. 

To express this implication as a probability, we first separate earthquake occurrence from earthquake size. 
Let $\lambda_0(t)$ be the rate above $M_c$ and $M_t\ge M_c$ the target magnitude. 
Multiplying this rate by the fraction of events exceeding $M_t$ gives the target rate, $\lambda_t(t)=\lambda_0(t)S(M_t\mid b(t),M_x(t))$. 

Then, the expected number of target events over a horizon $T$, denoted by $\Lambda_t(T)$, is
\begin{equation}
\Lambda_t(T)=\int_0^T\lambda_0(t)S(M_t\mid b(t),M_x(t))\,\mathrm dt\label{eq:maincount}
\end{equation}
Assuming a Poisson occurrence model with that rate path, the probability of zero target events within the same temporal horizon is $e^{-\Lambda_t(T)}$. 
Hence, the probability $P_t(T)$ of at least one target event, is its complement 
\begin{equation}
P_t(T)=1-e^{-\Lambda_t(T)}.\label{eq:mainprob}
\end{equation}
To understand the effect of time-varying interaction domains, let us define the taper suppression \(R(t)=[M_0(M_t)-M_0(M_c)]/M_0(M_x(t))\). This is the exponent that the taper contributes to the log-survivor at the target magnitude, \(S(M_t)=10^{-b(M_t-M_c)}\exp[-R(t)]\), so it measures in log units how strongly finite rupture capacity suppresses the target event. 
When the target lies well below the corner, \(R\) is small, the taper is negligible, and the target probability is governed by \(b\) and the event rate. 
When the target lies at or above the corner, \(R\) is of order unity or larger and the exponential factor controls the magnitude dependence.
A target well above the corner has large \(R\) and responds strongly to a rising corner; a target well below it responds weakly. 
Because \(R\) decreases as \(M_x\) rises, the sensitivity is largest while the corner lies just below the target and weakens once the corner has passed it. 
Supplementary equation~\eqref{eq:ratechange} separates the three contributions, with the corner term weighted by \(R\) and the slope term by the magnitude gap \(M_t-M_c\); their ratio measures which of the two magnitude-law changes dominates.

When the target is in the strongly tapered tail, raising the corner towards it changes the probability by many orders of magnitude. Once the corner exceeds the target, further growth has a smaller effect. Then, similar $b$-values can accompany very different large-event probabilities, and the corner can supply the dominant contribution to their change.

For a given target magnitude, the rate of earthquakes exceeding that magnitude is the product of the seismicity rate and the fraction of events large enough to exceed the target. At fixed \(b\), a higher \(M_x\) increases this fraction by weakening the taper. The expected number of large earthquakes over a time interval therefore depends on when \(M_x\) is high relative to when most earthquakes occur.
During background activity, the seismicity rate is approximately constant: if gradual loading raises \(M_x\), earthquakes occur at the same overall rate, but a growing fraction can reach the target magnitude. 
During foreshock acceleration, assuming accelerated moment release, great part of seismic activity of the sequence occurs during the late stage of the preparation process. If \(M_x\) also rises towards the mainshock, the two effects reinforce one another: more earthquakes occur per unit time, and each has a higher probability of reaching a large magnitude. 
After a mainshock, the timing is reversed. 
Aftershocks are mostly clustered just after the mainshock, so a high initial \(M_x\) has its greatest effect during the first burst of activity. If \(M_x\) then falls as stress relaxes, both the declining rate and the narrowing magnitude range reduce the expected number of large aftershocks. 
Stress transfer  may foster a high \(M_x\) on nearby faults, although the same corner reached later contributes fewer large events if activity has already declined\cite{dieterich1994,shcherbakov2004}.
Sections~\ref{sec:regimes} and~\ref{sec:probability} derive these three cases analytically.

\section*{Towards forecasting the changing probability of large earthquakes}
For a given seismicity rate, the probability of a large earthquake depends on the range of rupture sizes the fault system can generate. Our theory predicts that, as loading brings more patches close to failure, dynamic stress transfer can sustain instabilities over wider regions. 

In the magnitude distribution, this expansion shifts the corner magnitude \(M_x\) upwards, extending the range over which Gutenberg-Richter scaling applies. 
For a target magnitude above the background corner, the taper strongly suppresses the probability of reaching that magnitude. As \(M_x\) increases approaching that target, the model gives a much larger chances of occurrence to this part of the distribution. The resulting probability contrast can reach orders of magnitude at the same event rate and \(b\)-value.

The physical simulations show how this transition develops. As the applied stress increases, slip on one patch can trigger a cascade of failures over greater distances, hence involving larger areas. The ETAS simulations clarify how this process would appear in a catalog: at fixed \(b\)-value, an upward shift in \(M_x\) produces a shallower slope under an ordinary Gutenberg-Richter fit. Therefore, an apparent decrease in \(b\)-value assessed assuming pure power-law magnitude distribution may also mask an increasing size of the stress-interaction region. 

At Ridgecrest and L'Aquila, the curvature of the observed magnitude distributions provides evidence that the upper rupture scale changes. During foreshocks, estimates of \(M_x\) rise, but limited datasets sometimes challenge the estimation of its uncertainty. The decrease in the corner magnitude is clearly resolved during the aftershocks phase, which accounts for most of the information gain in the forecasts of future magnitudes (Supplementary Table~\ref{tab:scores}).

The challenge for future works and operational forecasting is to resolve the upward $M_x$ shift during loading, when a larger fraction of earthquakes may develop into major ruptures, but still few events have sampled that part of the distribution. 

Our framework and results prove that the taper of the earthquake frequency-magnitude distribution is an evolving property of the fault system. Fault dimensions constrain the maximum area available for rupture, while the stress level influences how much of that area can be activated at that time. 
As interaction regions expand or contract depending on the fault stress state, the corner magnitude \(M_x\), like the \(b\)-value, can change during the seismic cycle. 

The challenge for the future is to better resolve these shifts when few earthquakes sample the taper: measuring how the spatial correlations of seismicity evolve as well as how the magnitude distribution departs from a power law. The goal is to recognise, as a sequence evolves, when a large rupture is becoming more probable even though the seismicity rate and the \(b\)-value give little indication about that change.

\FloatBarrier\section*{Methods}
\subsection*{Earthquake selection and magnitude preprocessing}
We select earthquakes within 60\,km of each mainshock epicentre and at depths of 0-30\,km. 
For Ridgecrest, we utilize data from the Southern California Seismic Network catalogue\cite{scedc2013} from 6 July 2014 to 6 July 2021, centred on $35.7695^\circ$\,N, $117.59933^\circ$\,W. 
For L'Aquila, we analyse the Italian Seismological Instrumental and Parametric Database (ISIDe)\cite{iside} from 16 April 2005 to 6 April 2011, centred on $42.342^\circ$\,N, $13.380^\circ$\,E. The queries exclude the end date. The first selected L'Aquila event occurred on 19 April 2005. As a first step, we remove duplicate identifiers. 
We select the time zero as 03:19:53 UTC on 6 July 2019 at Ridgecrest and 01:32:40 UTC on 6 April 2009 at L'Aquila.
We use local magnitudes ($M_L$) rounded to 0.1. We exclude events lacking $M_L$. 
For descriptive frequency-magnitude distributions only, we replace the mainshock $M_L$ (6.7 at Ridgecrest and 5.9 at L'Aquila, which arise from magnitude saturation) with the estimated moment magnitude ($M_w$ 7.1 and 6.3, respectively\cite{ross2019,usgs2009laquila}).

\subsection*{Estimation of the slope, corner magnitude and uncertainty}
We calculated the Gutenberg-Richter (GR) $b$-value using the $b$-positive estimator\cite{vanderelst2021}, which uses the positive differences $m_i-m_{i-1}$ between the reported magnitudes of consecutive earthquakes in time order.

For the tapered Gutenberg-Richter (TGR) law\cite{kagan2002,bird2004}, we estimated jointly both the slope $b$ and the corner magnitude $M_x$ from the data. 
Let $\delta=0.1$ be the magnitude-bin width and $m_c$ the center of the lowest bin included in the fit. Its lower edge, $M_c=m_c-\delta/2$, defines the threshold of the continuous distribution. For a target magnitude $M\ge M_c$, the probability that an earthquake above this threshold reaches or exceeds $M$ is
\begin{equation}
	S(M)=10^{-b(M-M_c)}
	\exp\!\left\{-q\left[10^{1.5(M-M_c)}-1\right]\right\},
\end{equation}
where $q=10^{1.5(M_c-M_x)}$. Here, $b$ controls the slope below the taper, while $M_x$ sets the magnitude scale at which large earthquakes become less frequent than under the GR law. The parameter $q$ measures the taper strength. The factor 1.5 comes from the conversion relation between seismic moment and moment magnitude ($M_w$). 
For local magnitude ($M_L$), we use the same factor to describe the taper empirically\cite{hanks1979,baltay2025}.

Each magnitude $m$ represents the interval $[m-\delta/2,m+\delta/2)$. Therefore, we estimated $b$ and $q$ by maximising the log likelihood $\mathcal{L}$ of the observed magnitude bins,
\begin{equation}
	\mathcal{L}=\sum_i w_i\ln\pi(m_i),
	\label{eq:binprob}
\end{equation}
where $\pi(m) = S(m-\delta/2)-S(m+\delta/2)$ is the probability assigned to the bin centered on $m$, $m_i$ is the magnitude of earthquake $i$, and $w_i$ is its weight in the fit. For unweighted samples, $w_i=1$ for every earthquake.

We optimise the tapered fit comparing it with the GR solution at $q=0$. We profile each parameter by fixing its value and maximising $\mathcal{L}$ over the other\cite{schoenberg2001}. If an interval includes $q=0$, only its lower bound on $M_x$ is finite.
At each update, the weight $w_k$ determines how much an earthquake contributes to the log likelihood used to estimate $b$ and $M_x$. We give recent earthquakes greater influence so that the fit follows changes in the sequence. We introduce each event gradually to limit abrupt shifts in the estimated corner when a large earthquake first enters the sample.
To do this, we introduce weights as follows. 
Let $k$ count the earthquakes with $M_L\geq1.5$, so that $k=0$ stands by the most recent event. Among earthquakes above the completeness threshold for that update, we assign each event a fraction $p_k$ of the total weight, with
\[
p_k\propto e^{-k/200}\big[1-e^{-(k+1)/50}\big].
\]
Fractions are such that they sum to 1. The weight entering the likelihood is
\begin{equation}
	w_k=\frac{1}{\sum_j p_j^2}\,p_k,
	\label{eq:weights}
\end{equation}
If all $n$ events contribute equally, then $\frac{1}{\sum_j p_j^2}=n$ and each event has weight one. When a few events carry much of the weight, the pre-factor decreases. This scaling allows us to account for unequal earthquake contributions when we calculate their profile-likelihood intervals and to smooth fluctuations. 

For the $b$-value, we separate the time series into two parts by distinguishing seismicity before $M_w\,6.4$ and $M_w\,7.1$ Ridgecrest earthquakes and the $M_w\,6.3$ L'Aquila earthquake, and after. For each side, we rebuild the windows, $M_c$ estimates and event weights within each segment, as defined above. After each boundary, we resume plotting once the fitting and estimation criteria previously described are met. This procedure is applied in agreement with the procedures adopted in \cite{gulia2019} and works afterwards. 
Conversely, we use the continuous catalog for the corner-magnitude time series.

\subsection*{ETAS calibration and forecasting scores}
We use the epidemic-type aftershock sequence (ETAS) model to describe the rate of earthquakes with $m_j\ge m_c=2.0$\cite{ogata1988}. 
Given the earthquake process $\mathcal H_t$ before time $t$, the rate is
\begin{equation}
	\lambda_0(t\mid\mathcal H_t)=\mu_0+
	\sum_{t_j<t}K\,10^{\alpha(m_j-m_c)}\frac{(p-1)c^{p-1}}{(t-t_j+c)^p}.\label{eq:etas}
\end{equation}
The background rate is $\mu_0$ events per day. Each parent of magnitude $m_j$ at time $t_j$ produces $K10^{\alpha(m_j-m_c)}$ direct aftershocks on average. 
The coefficient $K$ is the productivity at $m_c$, $\alpha$ its magnitude dependence (called fertility exponent), $c$ the time offset in days and $p$ the exponent of aftershock decay. 
We fit the five parameters by maximising the log-likelihood of the point process over each training interval $[t_a,t_b)$,
\begin{equation}
	\mathcal{L}_{\rm time}=\sum_{t_a<t_i<t_b}\ln\lambda_0(t_i\mid\mathcal H_{t_i})
	-\int_{t_a}^{t_b}\lambda_0(t\mid\mathcal H_t)\,\mathrm dt .
\end{equation}
Training starts at the first event $t_a$ and ends on 1 January 2018 for the Ridgecrest or 1 January 2008 for L'Aquila seismicity. 

The rate in each magnitude bin is $\lambda_0(t)\pi_t(m)$, where $\pi_t(m)$ is its conditional probability. 
For each model, we calculate the Bayesian information criterion
\begin{equation}
   B=k\ln n-2\mathcal{L}.  
\end{equation}
Here $n$ is the number of earthquakes used to fit the GR and TGR models, $\mathcal{L}$ is the maximised magnitude log likelihood and $k$ is the number of degrees of freedom (one for GR and two for TGR)\cite{schwarz1978}. 
Then, the TGR weight is 
\begin{equation}
	w_T=\frac{1}{1+e^{(B_T-B_G)/2}},\label{eq:modelweight}
\end{equation}
and the probability to the next magnitude $m_i$ is 
\begin{equation}
	\pi_i(m_i)=w_T\pi_T(m_i)+(1-w_T)\pi_G(m_i),\label{eq:predictive}
\end{equation}
where $T$ and $G$ denote TGR and GR. We also score each component alone. Two further references use the same mixture fitted to the initial training period or to all the previous events.
For $N$ earthquakes, the cumulative information gain over a reference bin probability $\pi_{{\rm ref},i}$ is\cite{schorlemmer2007testing}
\begin{equation}
	G_N=\sum_{i=1}^{N}\ln\frac{\pi_i(m_i)}{\pi_{{\rm ref},i}(m_i)}.\label{eq:score}
\end{equation}

\subsection*{ETAS simulations}
We generate ETAS catalogs with predefined $b$ and $M_x$ to test whether the fits can  distinguish a changing slope from a moving corner. 
Following the same formulations of the previous section, background events are Poisson generated. To each earthquake a magnitude $m$ is assigned from the TGR distribution at its origin time; then we generate a Poisson number of aftershocks with mean corresponding to the productivity $K10^{\alpha(m-M_c)}$ and triggering times from the modified Omori law (equation~\eqref{eq:etas}). Aftershocks trigger further generations by the same rule. At each stage, we set $K=n/\mathbb E[10^{\alpha(m-M_c)}]$ to preserve the mean number of aftershocks per event, $n$ (branching ratio)\cite{helmstetter2002}.

We set $M_c=2$, $n=0.55$, $\alpha=0.5$, $c=0.01$ days and $p=1.2$ over $T = 2000$ days (corresponding to the duration of the Ridgecrest and L'Aquila selected datasets). 
We generate several catalogs with various combinations of $(b,M_x)$. 

At each stage, we fit magnitudes using the theoretical probability derived by the survivor function ($-\mathrm dS/\mathrm dM$) and estimate median and 2$\sigma$ uncertainties. We count how often the smallest corrected Akaike information criterion (AIC) choose the expected change\cite{hurvich1989}.

\subsection*{Physical simulations of heterogeneous faults}
We simulate the temporal evolution of the frequency-magnitude distribution on a rough, heterogeneous fault with realistic strength properties with focus on the $b$-value and the corner magnitude $M_x$ \cite{fisher1997,sornette1996}.

We represent the fault as a square plane divided into patches 100\,m on each side. To generate a fractal spatial pattern of strong and weak regions, we start from a two-dimensional Gaussian white-noise field and multiply each Fourier mode by $|\boldsymbol{k}|^{-0.6}$, where $\boldsymbol{k}$ is the spatial wave vector. 
We then transform the field back to physical space and standardize its values of $G_i$ to zero mean and unit variance. Since the white noise has a flat power spectrum and the power is proportional to the squared Fourier amplitude, the resulting spectrum scales as $|\boldsymbol{k}|^{-1.2}$. This field is characterized by larger weights for long wavelengths; at the same time, it shows variations even at rougher wavelengths, so that the strength varies over a range of distances along the fault.

Each patch $i$ has a fixed strength component, represented by $G_i$, and a component that changes when the patch slips. Let $h$ denote the local slip in arbitrary units, i.e., $\Delta u=1$\,mm. We obtain a failure strength $\tau_i(h)$ from
\begin{equation}
	\begin{cases}
		P_i(h)=\Phi\!\left(\sqrt{\rho}\,G_i+\sqrt{1-\rho}\,\eta_i(h)\right)\\[3pt]
		\tau_i(h)=\sqrt{-\ln\!\left[1-P_i(h)\right]}.
	\end{cases}
\end{equation}
We sample $\eta_i(h)$ from a standard normal distribution separately for each patch, first at initialisation and again after each local slip. The standard normal cumulative distribution $\Phi$ returns the strength cumulative value $P_i(h)$, which the second equation converts to a Weibull threshold. 
We choose the Weibull distribution because it is widely used to describe the variability of strength in brittle heterogeneous materials\cite{weibull1951}. The parameter $\rho$ determines the fraction of the Gaussian variance associated with the structural spatial pattern. We use $\rho=0.35$ for the simulations with correlated fault structure and $\rho=0$ for the comparison with spatially uncorrelated strengths.

At the beginning of the simulation, all patches start with zero slip. 
We apply a uniform stress load $F$; the resulting driving stress on patch $i$ is
\begin{equation}
	\sigma_i=F+k_e\sum_{j\in\mathcal N_i}(h_j-h_i),
\end{equation}
where $h_i$ is the accumulated slip on patch $i$ in $\Delta u$ units, $\mathcal N_i$ contains its four nearest neighbors and $k_e=0.2$ (different values do not alter the dynamics qualitatively) is the elastic coupling. 
We increase $F$ until the first patch reaches its failure strength, then hold $F$ fixed. At each step, all patches with $\sigma_i\geq\tau_i(h_i)$ slip by $\Delta u$ and acquire a new strength. One slip lowers the elastic stress on the slipping patch by $4k_e$ and raises it on each neighbor by $k_e$. Therefore, a rupture can spread throughout the fault at constant applied load (this is done to simulate the decoupled temporal scales of earthquakes and tectonic stress accumulation). Once all patches are stable, the earthquake ends and loading continues. The simulation stops after a fixed time $T$ depending on the required statistics. 

We calculate the seismic moment of each rupture from its slip distribution using 
\begin{equation}
	M_0=G a_0\sum_{i\in\mathcal R}d_i,
\end{equation}
where $G=30$\,GPa is the shear modulus, $a_0=(100\,\mathrm{m})^2$ is the patch area, $\mathcal R$ is the set of patches that slipped and $d_i$ is the total slip on patch $i$ during that earthquake. 
To test whether a rising $M_x$ produces an increase in the rupture area, we also calculate a moment from the number $N_A$ of distinct patches that slipped, employing 
\begin{equation}
	M_{0,A}=G a_0\Delta u\,N_A^{3/2}.
\end{equation}
The exponent $3/2$ follows from self-similar rupture at constant stress drop, for which mean slip scales with the square root of rupture area. Repeated slip on the same patch increases $M_0$, whereas $M_{0, A}$ increases only when the rupture reaches additional patches. For either moment expressed in N\,m, we calculate moment magnitude as $M_w=(2/3)(\log_{10}M_0-9.1)$\cite{hanks1979}.

To follow the magnitude distribution as the fault approaches failure, we collect ruptures by the fraction $F/F_f$ of the final load at which they occurred. 
Within each $F$ interval, we fit the GR and TGR laws using the probability of each rupture size. 
We repeat the analysis for different fault dimensions using both correlated and uncorrelated strengths to test how the rupture area and the spatial patterns of strength affect the evolution of $b$ and $M_x$. For the slip-based estimates, we obtain uncertainty intervals by resampling whole simulations, which keeps the ruptures from each loading sequence together. 

\subsection*{Probability of exceeding a target magnitude}
For a target magnitude $M_t$ and time horizon $T$, we compute the expected number of events using $\Lambda_t=\int_0^T\lambda_0(t)S(M_t\mid b(t),M_x(t))\,\mathrm dt$. Under the Poisson hypothesis, $1-e^{-\Lambda_t}$ gives the probability of at least one such earthquake. We vary $b$, $M_x$ or both for the same rate history. For background seismicity, foreshock acceleration (formulated via the accelerated seismic moment release) and aftershocks (modified Omori Law), we obtain $M_x(t)$ from the stress gap in equation~\eqref{eq:stresscorner} using the formulas in Supplementary Section~\ref{sec:regimes}.

\section*{Data availability}
The Ridgecrest catalog is available from the Southern California Earthquake Data Center\cite{scedc2013} (\url{https://scedc.caltech.edu/data/eq-catalogs.html}). The L'Aquila catalogue is available from the Istituto Nazionale di Geofisica e Vulcanologia through ISIDe\cite{iside} (\url{https://webservices.ingv.it/fdsnws/event/1/}). Map data come from ETOPO1, the California Geological Survey, the INGV--ISPRA database and Natural Earth\cite{amante2009,ponti2020,cinti2018database,naturalearthland}.

\section*{Code availability}
The MATLAB codes used in this study is available from the first author upon reasonable request.

\section*{Acknowledgments}
The authors thank Francesco Grigoli, Linxuan Li, Didier Sornette and Seth Stein for stimulating discussions. 
This research was partially supported by the European Research Council (ERC) Advanced Grant awarded to Laura Gulia for the project ``Forecasting and understanding the seIsmic cycle through b-value'' (FARIA, grant agreement no.~101200403).
The authors used DeepSeek to assist with improving the fluency of the manuscript and visualization, and perform code debugging. The authors take full responsibility for the analyses, their interpretation and the final content of the manuscript.

\section*{Author contributions}
D.Z. and L.G. designed the study and interpreted results together. D.Z. developed the theory and methods, realized the simulations, analyzed the data, prepared the figures and wrote the original manuscript. L.G. supervised the research, reviewed and edited the manuscript, provided funding.

\section*{Competing interests}
The authors declare no competing interests.

\setlength{\bibsep}{3pt}\bibliography{b_biblio}

\begin{thebibliography}{10}
\expandafter\ifx\csname url\endcsname\relax
  \def\url#1{\burl{#1}}\fi
\expandafter\ifx\csname urlprefix\endcsname\relax\def\urlprefix{URL }\fi
\providecommand{\bibinfo}[2]{#2}
\providecommand{\eprint}[2][]{\url{#2}}
\providecommand{\doi}[1]{\url{https://doi.org/#1}}
\bibcommenthead

\bibitem{gutenberg1944}
\bibinfo{author}{Gutenberg, B.} \& \bibinfo{author}{Richter, C.~F.}
\newblock \bibinfo{title}{{Frequency of earthquakes in California}}.
\newblock \emph{\bibinfo{journal}{Bulletin of the Seismological Society of
  America}} \textbf{\bibinfo{volume}{34}}, \bibinfo{pages}{185--188}
  (\bibinfo{year}{1944}).

\bibitem{ishimoto1939}
\bibinfo{author}{Ishimoto, M.} \& \bibinfo{author}{Iida, K.}
\newblock \bibinfo{title}{{Observations sur les s\'eismes enregistr\'es par le
  micro-s\'eismographe construit derni\`erement (1)}}.
\newblock \emph{\bibinfo{journal}{Bulletin of the Earthquake Research
  Institute, University of Tokyo}} \textbf{\bibinfo{volume}{17}},
  \bibinfo{pages}{443--478} (\bibinfo{year}{1939}).

\bibitem{kagan1999}
\bibinfo{author}{Kagan, Y.~Y.}
\newblock \bibinfo{title}{{Universality of the Seismic Moment-frequency
  Relation}}.
\newblock \emph{\bibinfo{journal}{Pure and Applied Geophysics}}
  \textbf{\bibinfo{volume}{155}}, \bibinfo{pages}{537--573}
  (\bibinfo{year}{1999}).

\bibitem{kagan2010}
\bibinfo{author}{Kagan, Y.~Y.}
\newblock \bibinfo{title}{{Earthquake size distribution: Power-law with
  exponent $\beta\equiv 1/2$?}}
\newblock \emph{\bibinfo{journal}{Tectonophysics}}
  \textbf{\bibinfo{volume}{490}}, \bibinfo{pages}{103--114}
  (\bibinfo{year}{2010}).

\bibitem{fisher1997}
\bibinfo{author}{Fisher, D.~S.}, \bibinfo{author}{Dahmen, K.},
  \bibinfo{author}{Ramanathan, S.} \& \bibinfo{author}{Ben-Zion, Y.}
\newblock \bibinfo{title}{{Statistics of Earthquakes in Simple Models of
  Heterogeneous Faults}}.
\newblock \emph{\bibinfo{journal}{Physical Review Letters}}
  \textbf{\bibinfo{volume}{78}}, \bibinfo{pages}{4885--4888}
  (\bibinfo{year}{1997}).

\bibitem{sethna2001}
\bibinfo{author}{Sethna, J.~P.}, \bibinfo{author}{Dahmen, K.~A.} \&
  \bibinfo{author}{Myers, C.~R.}
\newblock \bibinfo{title}{{Crackling noise}}.
\newblock \emph{\bibinfo{journal}{Nature}} \textbf{\bibinfo{volume}{410}},
  \bibinfo{pages}{242--250} (\bibinfo{year}{2001}).

\bibitem{wiemer1997}
\bibinfo{author}{Wiemer, S.} \& \bibinfo{author}{Wyss, M.}
\newblock \bibinfo{title}{{Mapping the frequency-magnitude distribution in
  asperities: An improved technique to calculate recurrence times?}}
\newblock \emph{\bibinfo{journal}{Journal of Geophysical Research: Solid
  Earth}} \textbf{\bibinfo{volume}{102}}, \bibinfo{pages}{15115--15128}
  (\bibinfo{year}{1997}).

\bibitem{schorlemmer2005}
\bibinfo{author}{Schorlemmer, D.}, \bibinfo{author}{Wiemer, S.} \&
  \bibinfo{author}{Wyss, M.}
\newblock \bibinfo{title}{{Variations in earthquake-size distribution across
  different stress regimes}}.
\newblock \emph{\bibinfo{journal}{Nature}} \textbf{\bibinfo{volume}{437}},
  \bibinfo{pages}{539--542} (\bibinfo{year}{2005}).

\bibitem{scholz2015}
\bibinfo{author}{Scholz, C.~H.}
\newblock \bibinfo{title}{{On the stress dependence of the earthquake b
  value}}.
\newblock \emph{\bibinfo{journal}{Geophysical Research Letters}}
  \textbf{\bibinfo{volume}{42}}, \bibinfo{pages}{1399--1402}
  (\bibinfo{year}{2015}).

\bibitem{scholz1968}
\bibinfo{author}{Scholz, C.~H.}
\newblock \bibinfo{title}{{The frequency-magnitude relation of microfracturing
  in rock and its relation to earthquakes}}.
\newblock \emph{\bibinfo{journal}{Bulletin of the Seismological Society of
  America}} \textbf{\bibinfo{volume}{58}}, \bibinfo{pages}{399--415}
  (\bibinfo{year}{1968}).

\bibitem{amitrano2003}
\bibinfo{author}{Amitrano, D.}
\newblock \bibinfo{title}{{Brittle-ductile transition and associated
  seismicity: Experimental and numerical studies and relationship with the b
  value}}.
\newblock \emph{\bibinfo{journal}{Journal of Geophysical Research: Solid
  Earth}} \textbf{\bibinfo{volume}{108}}, \bibinfo{pages}{2044}
  (\bibinfo{year}{2003}).

\bibitem{goebel2013}
\bibinfo{author}{Goebel, T. H.~W.}, \bibinfo{author}{Schorlemmer, D.},
  \bibinfo{author}{Becker, T.~W.}, \bibinfo{author}{Dresen, G.} \&
  \bibinfo{author}{Sammis, C.~G.}
\newblock \bibinfo{title}{{Acoustic emissions document stress changes over many
  seismic cycles in stick-slip experiments}}.
\newblock \emph{\bibinfo{journal}{Geophysical Research Letters}}
  \textbf{\bibinfo{volume}{40}}, \bibinfo{pages}{2049--2054}
  (\bibinfo{year}{2013}).

\bibitem{zaccagnino2022scaling}
\bibinfo{author}{Zaccagnino, D.}, \bibinfo{author}{Telesca, L.} \&
  \bibinfo{author}{Doglioni, C.}
\newblock \bibinfo{title}{{Scaling properties of seismicity and faulting}}.
\newblock \emph{\bibinfo{journal}{Earth and Planetary Science Letters}}
  \textbf{\bibinfo{volume}{584}}, \bibinfo{pages}{117511}
  (\bibinfo{year}{2022}).

\bibitem{nanjo2012}
\bibinfo{author}{Nanjo, K.~Z.}, \bibinfo{author}{Hirata, N.},
  \bibinfo{author}{Obara, K.} \& \bibinfo{author}{Kasahara, K.}
\newblock \bibinfo{title}{{Decade-scale decrease in b value prior to the
  M9-class 2011 Tohoku and 2004 Sumatra quakes}}.
\newblock \emph{\bibinfo{journal}{Geophysical Research Letters}}
  \textbf{\bibinfo{volume}{39}}, \bibinfo{pages}{L20304}
  (\bibinfo{year}{2012}).

\bibitem{gulia2010}
\bibinfo{author}{Gulia, L.} \& \bibinfo{author}{Wiemer, S.}
\newblock \bibinfo{title}{{The influence of tectonic regimes on the earthquake
  size distribution: A case study for Italy}}.
\newblock \emph{\bibinfo{journal}{Geophysical Research Letters}}
  \textbf{\bibinfo{volume}{37}}, \bibinfo{pages}{L10305}
  (\bibinfo{year}{2010}).

\bibitem{gulia2020}
\bibinfo{author}{Gulia, L.}, \bibinfo{author}{Wiemer, S.} \&
  \bibinfo{author}{Vannucci, G.}
\newblock \bibinfo{title}{{Pseudoprospective Evaluation of the Foreshock
  Traffic-Light System in Ridgecrest and Implications for Aftershock Hazard
  Assessment}}.
\newblock \emph{\bibinfo{journal}{Seismological Research Letters}}
  \textbf{\bibinfo{volume}{91}}, \bibinfo{pages}{2828--2842}
  (\bibinfo{year}{2020}).

\bibitem{gulia2019}
\bibinfo{author}{Gulia, L.} \& \bibinfo{author}{Wiemer, S.}
\newblock \bibinfo{title}{{Real-time discrimination of earthquake foreshocks
  and aftershocks}}.
\newblock \emph{\bibinfo{journal}{Nature}} \textbf{\bibinfo{volume}{574}},
  \bibinfo{pages}{193--199} (\bibinfo{year}{2019}).

\bibitem{mignan2014debate}
\bibinfo{author}{Mignan, A.}
\newblock \bibinfo{title}{The debate on the prognostic value of earthquake
  foreshocks: A meta-analysis}.
\newblock \emph{\bibinfo{journal}{Scientific Reports}}
  \textbf{\bibinfo{volume}{4}}, \bibinfo{pages}{4099} (\bibinfo{year}{2014}).

\bibitem{benzion2008}
\bibinfo{author}{Ben-Zion, Y.}
\newblock \bibinfo{title}{{Collective behavior of earthquakes and faults:
  Continuum-discrete transitions, progressive evolutionary changes, and
  different dynamic regimes}}.
\newblock \emph{\bibinfo{journal}{Reviews of Geophysics}}
  \textbf{\bibinfo{volume}{46}}, \bibinfo{pages}{RG4006}
  (\bibinfo{year}{2008}).

\bibitem{kato2021}
\bibinfo{author}{Kato, A.} \& \bibinfo{author}{Ben-Zion, Y.}
\newblock \bibinfo{title}{{The generation of large earthquakes}}.
\newblock \emph{\bibinfo{journal}{Nature Reviews Earth \& Environment}}
  \textbf{\bibinfo{volume}{2}}, \bibinfo{pages}{26--39} (\bibinfo{year}{2021}).

\bibitem{zaccagnino2021stress}
\bibinfo{author}{Zaccagnino, D.}, \bibinfo{author}{Telesca, L.} \&
  \bibinfo{author}{Doglioni, C.}
\newblock \bibinfo{title}{{Different Fault Response to Stress during the
  Seismic Cycle}}.
\newblock \emph{\bibinfo{journal}{Applied Sciences}}
  \textbf{\bibinfo{volume}{11}}, \bibinfo{pages}{9596} (\bibinfo{year}{2021}).

\bibitem{brodsky2014triggering}
\bibinfo{author}{Brodsky, E.~E.} \& \bibinfo{author}{van~der Elst, N.~J.}
\newblock \bibinfo{title}{{The Uses of Dynamic Earthquake Triggering}}.
\newblock \emph{\bibinfo{journal}{Annual Review of Earth and Planetary
  Sciences}} \textbf{\bibinfo{volume}{42}}, \bibinfo{pages}{317--339}
  (\bibinfo{year}{2014}).

\bibitem{zaccagnino2024foreshocks}
\bibinfo{author}{Zaccagnino, D.}, \bibinfo{author}{Vallianatos, F.},
  \bibinfo{author}{Michas, G.}, \bibinfo{author}{Telesca, L.} \&
  \bibinfo{author}{Doglioni, C.}
\newblock \bibinfo{title}{{Are Foreshocks Fore-Shocks?}}
\newblock \emph{\bibinfo{journal}{Journal of Geophysical Research: Solid
  Earth}} \textbf{\bibinfo{volume}{129}}, \bibinfo{pages}{e2023JB027337}
  (\bibinfo{year}{2024}).

\bibitem{zaccagnino2026foreshocks}
\bibinfo{author}{Zaccagnino, D.} \& \bibinfo{author}{Doglioni, C.}
\newblock \bibinfo{title}{{A statistical approach to foreshocks discrimination
  in Italy}}.
\newblock \emph{\bibinfo{journal}{Natural Hazards}}
  \textbf{\bibinfo{volume}{122}}, \bibinfo{pages}{84} (\bibinfo{year}{2026}).

\bibitem{zoller2001correlation}
\bibinfo{author}{Z{\"o}ller, G.}, \bibinfo{author}{Hainzl, S.} \&
  \bibinfo{author}{Kurths, J.}
\newblock \bibinfo{title}{Observation of growing correlation length as an
  indicator for critical point behavior prior to large earthquakes}.
\newblock \emph{\bibinfo{journal}{Journal of Geophysical Research: Solid
  Earth}} \textbf{\bibinfo{volume}{106}}, \bibinfo{pages}{2167--2175}
  (\bibinfo{year}{2001}).

\bibitem{zoller2002critical}
\bibinfo{author}{Z{\"o}ller, G.} \& \bibinfo{author}{Hainzl, S.}
\newblock \bibinfo{title}{A systematic spatiotemporal test of the critical
  point hypothesis for large earthquakes}.
\newblock \emph{\bibinfo{journal}{Geophysical Research Letters}}
  \textbf{\bibinfo{volume}{29}}, \bibinfo{pages}{53--1--53--4}
  (\bibinfo{year}{2002}).

\bibitem{benzion2003criticality}
\bibinfo{author}{Ben-Zion, Y.}, \bibinfo{author}{Eneva, M.} \&
  \bibinfo{author}{Liu, Y.}
\newblock \bibinfo{title}{Large earthquake cycles and intermittent criticality
  on heterogeneous faults due to evolving stress and seismicity}.
\newblock \emph{\bibinfo{journal}{Journal of Geophysical Research: Solid
  Earth}} \textbf{\bibinfo{volume}{108}}, \bibinfo{pages}{2307}
  (\bibinfo{year}{2003}).

\bibitem{zaccagnino2025santorini}
\bibinfo{author}{Zaccagnino, D.}, \bibinfo{author}{Michas, G.},
  \bibinfo{author}{Telesca, L.} \& \bibinfo{author}{Vallianatos, F.}
\newblock \bibinfo{title}{{Precursory patterns, evolution and physical
  interpretation of the 2025 Santorini-Amorgos seismic sequence}}.
\newblock \emph{\bibinfo{journal}{Earth and Planetary Science Letters}}
  \textbf{\bibinfo{volume}{671}}, \bibinfo{pages}{119656}
  (\bibinfo{year}{2025}).

\bibitem{roche2026probability}
\bibinfo{author}{Roche, V.}, \bibinfo{author}{van~der Baan, M.},
  \bibinfo{author}{Walsh, J.} \& \bibinfo{author}{Manzocchi, T.}
\newblock \bibinfo{title}{Probability and recurrence of maximum magnitude based
  on stochastic modeling of internal fault geometry}.
\newblock \emph{\bibinfo{journal}{Journal of Geophysical Research: Solid
  Earth}} \textbf{\bibinfo{volume}{131}}, \bibinfo{pages}{e2025JB032073}
  (\bibinfo{year}{2026}).

\bibitem{rodriguez2026role}
\bibinfo{author}{Rodriguez~Piceda, C.} \emph{et~al.}
\newblock \bibinfo{title}{The role of fault network geometry on the complexity
  of seismic cycles in the apennines}.
\newblock \emph{\bibinfo{journal}{Solid Earth}} \textbf{\bibinfo{volume}{17}},
  \bibinfo{pages}{311--346} (\bibinfo{year}{2026}).

\bibitem{kagan2002}
\bibinfo{author}{Kagan, Y.~Y.}
\newblock \bibinfo{title}{{Seismic moment distribution revisited: I.
  Statistical results}}.
\newblock \emph{\bibinfo{journal}{Geophysical Journal International}}
  \textbf{\bibinfo{volume}{148}}, \bibinfo{pages}{520--541}
  (\bibinfo{year}{2002}).

\bibitem{bird2004}
\bibinfo{author}{Bird, P.} \& \bibinfo{author}{Kagan, Y.~Y.}
\newblock \bibinfo{title}{{Plate-Tectonic Analysis of Shallow Seismicity:
  Apparent Boundary Width, Beta, Corner Magnitude, Coupled Lithosphere
  Thickness, and Coupling in Seven Tectonic Settings}}.
\newblock \emph{\bibinfo{journal}{Bulletin of the Seismological Society of
  America}} \textbf{\bibinfo{volume}{94}}, \bibinfo{pages}{2380--2399}
  (\bibinfo{year}{2004}).

\bibitem{hirata1989}
\bibinfo{author}{Hirata, T.}
\newblock \bibinfo{title}{{A correlation between the b value and the fractal
  dimension of earthquakes}}.
\newblock \emph{\bibinfo{journal}{Journal of Geophysical Research: Solid
  Earth}} \textbf{\bibinfo{volume}{94}}, \bibinfo{pages}{7507--7514}
  (\bibinfo{year}{1989}).

\bibitem{hanks1979}
\bibinfo{author}{Hanks, T.~C.} \& \bibinfo{author}{Kanamori, H.}
\newblock \bibinfo{title}{{A moment magnitude scale}}.
\newblock \emph{\bibinfo{journal}{Journal of Geophysical Research: Solid
  Earth}} \textbf{\bibinfo{volume}{84}}, \bibinfo{pages}{2348--2350}
  (\bibinfo{year}{1979}).

\bibitem{aki1981}
\bibinfo{author}{Aki, K.}
\newblock \bibinfo{title}{ in \textit{{A Probabilistic Synthesis of Precursory
  Phenomena}}} (eds \bibinfo{editor}{Simpson, D.~W.} \&
  \bibinfo{editor}{Richards, P.~G.}) \emph{\bibinfo{booktitle}{Earthquake
  Prediction: An International Review}}, Vol.~\bibinfo{volume}{4} of
  \emph{\bibinfo{series}{Maurice Ewing Series}} \bibinfo{pages}{566--574}
  (\bibinfo{publisher}{American Geophysical Union}, \bibinfo{year}{1981}).

\bibitem{ledoussal2009}
\bibinfo{author}{Le~Doussal, P.} \& \bibinfo{author}{Wiese, K.~J.}
\newblock \bibinfo{title}{{Size distributions of shocks and static avalanches
  from the functional renormalization group}}.
\newblock \emph{\bibinfo{journal}{Physical Review E}}
  \textbf{\bibinfo{volume}{79}}, \bibinfo{pages}{051106}
  (\bibinfo{year}{2009}).

\bibitem{rosso2009}
\bibinfo{author}{Rosso, A.}, \bibinfo{author}{Le~Doussal, P.} \&
  \bibinfo{author}{Wiese, K.~J.}
\newblock \bibinfo{title}{{Avalanche-size distribution at the depinning
  transition: A numerical test of the theory}}.
\newblock \emph{\bibinfo{journal}{Physical Review B}}
  \textbf{\bibinfo{volume}{80}}, \bibinfo{pages}{144204}
  (\bibinfo{year}{2009}).

\bibitem{ross2019}
\bibinfo{author}{Ross, Z.~E.} \emph{et~al.}
\newblock \bibinfo{title}{{Hierarchical interlocked orthogonal faulting in the
  2019 Ridgecrest earthquake sequence}}.
\newblock \emph{\bibinfo{journal}{Science}} \textbf{\bibinfo{volume}{366}},
  \bibinfo{pages}{346--351} (\bibinfo{year}{2019}).

\bibitem{atterholt2025}
\bibinfo{author}{Atterholt, J.}, \bibinfo{author}{Wilding, J.~D.} \&
  \bibinfo{author}{Ross, Z.~E.}
\newblock \bibinfo{title}{{The evolution of fault orientation in the 2019
  Ridgecrest earthquake sequence with a new long-term catalogue of seismicity
  and moment tensors}}.
\newblock \emph{\bibinfo{journal}{Geophysical Journal International}}
  \textbf{\bibinfo{volume}{240}}, \bibinfo{pages}{1579--1592}
  (\bibinfo{year}{2025}).

\bibitem{schorlemmer2007testing}
\bibinfo{author}{Schorlemmer, D.}, \bibinfo{author}{Gerstenberger, M.~C.},
  \bibinfo{author}{Wiemer, S.}, \bibinfo{author}{Jackson, D.~D.} \&
  \bibinfo{author}{Rhoades, D.~A.}
\newblock \bibinfo{title}{Earthquake likelihood model testing}.
\newblock \emph{\bibinfo{journal}{Seismological Research Letters}}
  \textbf{\bibinfo{volume}{78}}, \bibinfo{pages}{17--29}
  (\bibinfo{year}{2007}).

\bibitem{ogata1988}
\bibinfo{author}{Ogata, Y.}
\newblock \bibinfo{title}{{Statistical Models for Earthquake Occurrences and
  Residual Analysis for Point Processes}}.
\newblock \emph{\bibinfo{journal}{Journal of the American Statistical
  Association}} \textbf{\bibinfo{volume}{83}}, \bibinfo{pages}{9--27}
  (\bibinfo{year}{1988}).

\bibitem{werner2011}
\bibinfo{author}{Werner, M.~J.}, \bibinfo{author}{Helmstetter, A.},
  \bibinfo{author}{Jackson, D.~D.} \& \bibinfo{author}{Kagan, Y.~Y.}
\newblock \bibinfo{title}{High-resolution long-term and short-term earthquake
  forecasts for {California}}.
\newblock \emph{\bibinfo{journal}{Bulletin of the Seismological Society of
  America}} \textbf{\bibinfo{volume}{101}}, \bibinfo{pages}{1630--1648}
  (\bibinfo{year}{2011}).

\bibitem{ponti2020}
\bibinfo{author}{Ponti, D.~J.} \emph{et~al.}
\newblock \bibinfo{title}{{Documentation of Surface Fault Rupture and
  Ground-Deformation Features Produced by the 4 and 5 July 2019
  $M_{\mathrm{w}}$ 6.4 and $M_{\mathrm{w}}$ 7.1 Ridgecrest Earthquake
  Sequence}}.
\newblock \emph{\bibinfo{journal}{Seismological Research Letters}}
  \textbf{\bibinfo{volume}{91}}, \bibinfo{pages}{2942--2959}
  (\bibinfo{year}{2020}).

\bibitem{helmstetter2003}
\bibinfo{author}{Helmstetter, A.}, \bibinfo{author}{Sornette, D.} \&
  \bibinfo{author}{Grasso, J.-R.}
\newblock \bibinfo{title}{{Mainshocks are aftershocks of conditional
  foreshocks: How do foreshock statistical properties emerge from aftershock
  laws}}.
\newblock \emph{\bibinfo{journal}{Journal of Geophysical Research: Solid
  Earth}} \textbf{\bibinfo{volume}{108}}, \bibinfo{pages}{2046}
  (\bibinfo{year}{2003}).

\bibitem{dieterich1994}
\bibinfo{author}{Dieterich, J.}
\newblock \bibinfo{title}{{A constitutive law for rate of earthquake production
  and its application to earthquake clustering}}.
\newblock \emph{\bibinfo{journal}{Journal of Geophysical Research: Solid
  Earth}} \textbf{\bibinfo{volume}{99}}, \bibinfo{pages}{2601--2618}
  (\bibinfo{year}{1994}).

\bibitem{shcherbakov2004}
\bibinfo{author}{Shcherbakov, R.}, \bibinfo{author}{Turcotte, D.~L.} \&
  \bibinfo{author}{Rundle, J.~B.}
\newblock \bibinfo{title}{{A generalized Omori's law for earthquake aftershock
  decay}}.
\newblock \emph{\bibinfo{journal}{Geophysical Research Letters}}
  \textbf{\bibinfo{volume}{31}}, \bibinfo{pages}{L11613}
  (\bibinfo{year}{2004}).

\bibitem{scedc2013}
\bibinfo{author}{{Southern California Earthquake Data Center}}.
\newblock \bibinfo{title}{{Southern California Earthquake Data Center}}.
\newblock \bibinfo{howpublished}{Caltech} (\bibinfo{year}{2013}).
\newblock \urlprefix\url{https://scedc.caltech.edu/data/eq-catalogs.html}.
\newblock \bibinfo{note}{Catalog snapshot retrieved 29 October 2026}.

\bibitem{iside}
\bibinfo{author}{{ISIDe Working Group}}.
\newblock \bibinfo{title}{{Italian Seismological Instrumental and Parametric
  Database (ISIDe)}}.
\newblock \bibinfo{howpublished}{Istituto Nazionale di Geofisica e Vulcanologia
  (INGV)} (\bibinfo{year}{2007}).
\newblock \urlprefix\url{https://webservices.ingv.it/fdsnws/event/1/}.
\newblock \bibinfo{note}{Catalog snapshot retrieved 29 October 2026}.

\bibitem{usgs2009laquila}
\bibinfo{author}{{U.S. Geological Survey}}.
\newblock \bibinfo{title}{{M 6.3 -- 3 km SE of Sassa, Italy: 6 April 2009}}.
\newblock
  \bibinfo{howpublished}{\url{https://earthquake.usgs.gov/earthquakes/eventpage/usp000gvtu}}
  (\bibinfo{year}{2009}).
\newblock \bibinfo{note}{Reviewed event record; accessed 29 October 2026}.

\bibitem{vanderelst2021}
\bibinfo{author}{van~der Elst, N.~J.}
\newblock \bibinfo{title}{{B-Positive: A Robust Estimator of Aftershock
  Magnitude Distribution in Transiently Incomplete Catalogs}}.
\newblock \emph{\bibinfo{journal}{Journal of Geophysical Research: Solid
  Earth}} \textbf{\bibinfo{volume}{126}}, \bibinfo{pages}{e2020JB021027}
  (\bibinfo{year}{2021}).

\bibitem{baltay2025}
\bibinfo{author}{Baltay, A.} \& \bibinfo{author}{Abercrombie, R.~E.}
\newblock \bibinfo{title}{{Seismic Moment and Local Magnitude Scales in
  Ridgecrest, California, from the SCEC/USGS Community Stress Drop Validation
  Study}}.
\newblock \emph{\bibinfo{journal}{Bulletin of the Seismological Society of
  America}} \textbf{\bibinfo{volume}{115}}, \bibinfo{pages}{1279--1293}
  (\bibinfo{year}{2025}).

\bibitem{schoenberg2001}
\bibinfo{author}{Kagan, Y.~Y.} \& \bibinfo{author}{Schoenberg, F.~P.}
\newblock \bibinfo{title}{{Estimation of the upper cutoff parameter for the
  tapered Pareto distribution}}.
\newblock \emph{\bibinfo{journal}{Journal of Applied Probability}}
  \textbf{\bibinfo{volume}{38}}, \bibinfo{pages}{158--175}
  (\bibinfo{year}{2001}).

\bibitem{schwarz1978}
\bibinfo{author}{Schwarz, G.}
\newblock \bibinfo{title}{{Estimating the Dimension of a Model}}.
\newblock \emph{\bibinfo{journal}{The Annals of Statistics}}
  \textbf{\bibinfo{volume}{6}}, \bibinfo{pages}{461--464}
  (\bibinfo{year}{1978}).

\bibitem{helmstetter2002}
\bibinfo{author}{Helmstetter, A.} \& \bibinfo{author}{Sornette, D.}
\newblock \bibinfo{title}{{Subcritical and supercritical regimes in epidemic
  models of earthquake aftershocks}}.
\newblock \emph{\bibinfo{journal}{Journal of Geophysical Research: Solid
  Earth}} \textbf{\bibinfo{volume}{107}}, \bibinfo{pages}{2237}
  (\bibinfo{year}{2002}).

\bibitem{hurvich1989}
\bibinfo{author}{Hurvich, C.~M.} \& \bibinfo{author}{Tsai, C.-L.}
\newblock \bibinfo{title}{{Regression and time series model selection in small
  samples}}.
\newblock \emph{\bibinfo{journal}{Biometrika}} \textbf{\bibinfo{volume}{76}},
  \bibinfo{pages}{297--307} (\bibinfo{year}{1989}).

\bibitem{sornette1996}
\bibinfo{author}{Sornette, D.} \& \bibinfo{author}{Vanneste, C.}
\newblock \bibinfo{title}{{Faults self-organized by repeated earthquakes in a
  quasi-static antiplane crack model}}.
\newblock \emph{\bibinfo{journal}{Nonlinear Processes in Geophysics}}
  \textbf{\bibinfo{volume}{3}}, \bibinfo{pages}{1--12} (\bibinfo{year}{1996}).

\bibitem{weibull1951}
\bibinfo{author}{Weibull, W.}
\newblock \bibinfo{title}{{A Statistical Distribution Function of Wide
  Applicability}}.
\newblock \emph{\bibinfo{journal}{Journal of Applied Mechanics}}
  \textbf{\bibinfo{volume}{18}}, \bibinfo{pages}{293--297}
  (\bibinfo{year}{1951}).

\bibitem{amante2009}
\bibinfo{author}{Amante, C.} \& \bibinfo{author}{Eakins, B.~W.}
\newblock \bibinfo{title}{{{ETOPO1} 1 Arc-Minute Global Relief Model:
  Procedures, Data Sources and Analysis}}.
\newblock \bibinfo{type}{Technical Memorandum NESDIS NGDC}
  \bibinfo{number}{24}, \bibinfo{institution}{NOAA} (\bibinfo{year}{2009}).

\bibitem{cinti2018database}
\bibinfo{author}{Cinti, F.~R.}, \bibinfo{author}{Civico, R.},
  \bibinfo{author}{Blumetti, A.~M.}, \bibinfo{author}{Guerrieri, L.} \&
  \bibinfo{author}{Leoni, G.}
\newblock \bibinfo{title}{{INGV - ISPRA joint Surface Faulting Database - Mw
  6.1, 2009, April 6th L'Aquila earthquake (Central Italy)}}.
\newblock \bibinfo{howpublished}{PANGAEA} (\bibinfo{year}{2018}).
\newblock \bibinfo{note}{Dataset, CC-BY-4.0}.

\bibitem{naturalearthland}
\bibinfo{author}{{Natural Earth}}.
\newblock \bibinfo{title}{{1:110m Land polygons}}.
\newblock
  \bibinfo{howpublished}{\url{https://www.naturalearthdata.com/downloads/110m-physical-vectors/110m-land/}}
  (\bibinfo{year}{n.d.}).
\newblock \bibinfo{note}{Public-domain cartographic data; accessed 29 October
  2026}.

\bibitem{aki1965}
\bibinfo{author}{Aki, K.}
\newblock \bibinfo{title}{{Maximum likelihood estimate of b in the formula log
  N = a - b M and its confidence limits}}.
\newblock \emph{\bibinfo{journal}{Bulletin of the Earthquake Research
  Institute, University of Tokyo}} \textbf{\bibinfo{volume}{43}},
  \bibinfo{pages}{237--239} (\bibinfo{year}{1965}).

\bibitem{li2026distinguishing}
\bibinfo{author}{Li, L.} \& \bibinfo{author}{Avouac, J.-P.}
\newblock \bibinfo{title}{Distinguishing tapered and non-tapered
  {Gutenberg--Richter} distributions}.
\newblock \emph{\bibinfo{journal}{Bulletin of the Seismological Society of
  America}} \textbf{\bibinfo{volume}{116}}, \bibinfo{pages}{436--466}
  (\bibinfo{year}{2026}).

\bibitem{stauffer1994}
\bibinfo{author}{Stauffer, D.} \& \bibinfo{author}{Aharony, A.}
\newblock \emph{\bibinfo{title}{{Introduction to Percolation Theory}}}
  \bibinfo{edition}{2nd, revised} edn (\bibinfo{publisher}{Taylor \& Francis},
  \bibinfo{year}{1994}).

\bibitem{weinrib1983}
\bibinfo{author}{Weinrib, A.} \& \bibinfo{author}{Halperin, B.~I.}
\newblock \bibinfo{title}{{Critical phenomena in systems with
  long-range-correlated quenched disorder}}.
\newblock \emph{\bibinfo{journal}{Physical Review B}}
  \textbf{\bibinfo{volume}{27}}, \bibinfo{pages}{413--427}
  (\bibinfo{year}{1983}).

\end{thebibliography}

\clearpage
\begin{center}
	\section*{SUPPLEMENTARY INFORMATION}
\end{center}
\setcounter{section}{0}\setcounter{equation}{0}\setcounter{figure}{0}\setcounter{table}{0}\renewcommand{\theHequation}{supp.\arabic{equation}}\renewcommand{\theHfigure}{supp.\arabic{figure}}\renewcommand{\theHtable}{supp.\arabic{table}}
\renewcommand{\thesection}{S\arabic{section}}\renewcommand{\theequation}{S\arabic{equation}}\renewcommand{\thefigure}{S\arabic{figure}}\renewcommand{\thetable}{S\arabic{table}}

\section{Bias in the b-value estimation due to physical tapering}\label{sec:bias}
Fitting a tapered frequency-magnitude distribution with a power law introduces an upward bias in the b-value magnitude estimation. 
For a tapered Gutenberg-Richter (TGR) distribution, the probability that an earthquake above the threshold $M_c$ reaches magnitude $M$ is\cite{kagan2002,bird2004}
\begin{equation}
 S(M)=10^{-b(M-M_c)}
 \exp\!\left[-\left(10^{1.5(M-M_x)}-10^{1.5(M_c-M_x)}\right)\right].
 \label{eq:expsurvivor}
\end{equation}
Here, $M_x$ is the corner magnitude, and the expression applies at and above $M_c$ (equation~\eqref{eq:mainTGR}).

Suppose we fit these magnitudes with a GR distribution. Its log likelihood for $n$ magnitudes $M_i$ is\cite{aki1965}
\begin{equation*}
 \mathcal{L}(b_{\rm GR})=n\ln\!\big[b_{\rm GR}\ln(10)\big]
 -b_{\rm GR}\ln(10)\sum_{i=1}^n(M_i-M_c).
\end{equation*}
Differentiating with respect to the slope gives $\mathcal{L}'(b_{\rm GR})=n/b_{\rm GR}-\ln(10)\sum_{i=1}^n(M_i-M_c)$. Setting this derivative to zero gives the maximum-likelihood estimate
\begin{equation}
 \widehat b_{\rm GR}=\frac{n}{\ln(10)\sum_{i=1}^n(M_i-M_c)}.
 \label{eq:expectedgr}
\end{equation}
As the sample grows, the average of the observed differences $M_i-M_c$ approaches their expectation, i.e., the area under the survivor curve,
\begin{equation}
 \mathbb E[M-M_c]=\int_{M_c}^{\infty}S(M)\,\mathrm dM .
 \label{eq:meanexcess}
\end{equation}
For a finite corner, the exponential factor in equation~\eqref{eq:expsurvivor} is less than one at every $M>M_c$. Therefore, the integral is smaller than its expected GR value, $1/[b\ln(10)]$. 
Equation~\eqref{eq:expectedgr} then gives a large-sample GR slope greater than the theoretical $b$. Raising $M_x$ weakens the taper and increases the mean, so the fitted GR slope falls towards $b$ (Fig.~\ref{fig:s6}A).

The same effect is visible in the local slope of the exceedance curve. Differentiating equation~\eqref{eq:expsurvivor} gives
\begin{equation}
 -\frac{\mathrm d\log_{10}S(M)}{\mathrm dM}=b+1.5\times 10^{1.5(M-M_x)}.
 \label{eq:localslope}
\end{equation}
The curve steepens as $M$ approaches the corner (Fig.~\ref{fig:s6}B). Equations~\eqref{eq:expectedgr} and~\eqref{eq:meanexcess} describe the ordinary GR fit in this comparison. We use the $b$-positive method to estimate the $b$-value\cite{vanderelst2021}.

Therefore, it is crucial to correctly distinguish between tapered and standard Gutenberg-Richter magnitude distributions \cite{li2026distinguishing} before estimating the b-value, especially when the range of available magnitudes is tight. 

As an example, in the ETAS simulations, raising $M_x$ from 3 to 4.2 at fixed $b=1$ lowers the median GR estimate of about 20\% (Fig.~\ref{fig:s5}).

\paperfig{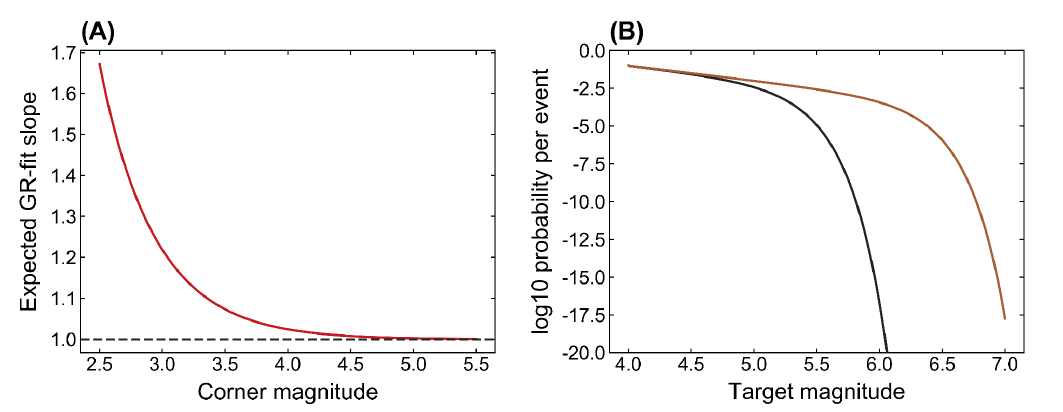}{\textbf{Effect of physical tapering on the slope of the Gutenberg-Richter law and the upper tail.} \textbf{A)} GR slope for TGR-distributed magnitudes above $M_c$ and $b=1$, calculated from equations~\eqref{eq:expectedgr} and~\eqref{eq:meanexcess}. \textbf{B)} Probability that one event above $M_c$ reaches a given magnitude, for $M_x=5$ (dark brown) and $M_x=6$ (copper) at $b=1$.}{fig:s6}

\section{How stress changes the \texorpdfstring{$b$}{b}-value}\label{sec:stress}
Stress can change the slope of the Gutenberg-Richter law by modulating the relative abundance of long ruptures or the slip they accumulate\cite{scholz2015,goebel2013}. At a given time, let $N(\geq L)\propto L^{-D}$ be the number of ruptures of length at least $L$, and let their seismic moment scale as $M_0\propto L^\rho$. Here, $D$ describes the distribution of lengths and $\rho$ describes how moment increases with length. Substituting $L\propto M_0^{1/\rho}$ into the length distribution gives $N(\geq M_0)\propto M_0^{-D/\rho}$. The moment--magnitude relation then gives $N(\geq M_w)\propto 10^{-[3D/(2\rho)]M_w}$. 
Comparing the exponent with the GR law gives\cite{aki1981,hanks1979}
\begin{equation}
	b=\frac{3D}{2\rho}.
	\label{eq:lengthb}
\end{equation}
If increasing stress favors long ruptures, $D$ decreases. If slip grows more rapidly with rupture length for the same rupture geometry, $\rho$ increases. Both changes lower $b$ when the other exponent is fixed.

The same calculation can be performed using the rupture area $A$ as variable. If its probability density scales as $A^{-\tau_A}$ with $\tau_A>1$, the number of ruptures with area at least $A$ is $\sim A^{1-\tau_A}$, if in the power-law range. Substituting $A\propto M_0^{1/\gamma}$ gives $N(\geq M_0)\propto M_0^{-(\tau_A-1)/\gamma}$, where $\gamma$ is the moment-area exponent. Converting moment to magnitude gives
\begin{equation}
	b=\frac{3(\tau_A-1)}{2\gamma}.
	\label{eq:areab}
\end{equation}
If we assume self-similar ruptures, $A\propto L^2$, so $D=2(\tau_A-1)$ and $\rho=2\gamma$.

To express the length distribution in terms of rupture propagation, consider all ruptures that reach length $L$ at a given value of shear stress $\Sigma$. Let us call $P_r(\Sigma)$ the fraction that reach at least $rL$, with $r>1$. This fraction is the ratio of the cumulative rupture number at $rL$ and $L$. Assuming that both the length from $L$ to $rL$ belong to the power-law range, the dependence on the starting length cancels, giving
\begin{equation}
	P_r(\Sigma)=r^{-D(\Sigma)}.
\end{equation}
Therefore, a smaller $D$ means that a greater fraction of ruptures reach any given multiple of their starting length. Substituting $D(\Sigma)$ into equation~\eqref{eq:lengthb} gives
\begin{equation}
	b(\Sigma)=-\frac{3}{2\rho}\frac{\ln P_r(\Sigma)}{\ln r}.
	\label{eq:stressb}
\end{equation}
This expression gives the same $b$ for any choice of $r$ within the scaling range. For self-similar ruptures with constant stress drop, $\rho=3$ and $b=D/2$.

To describe how loading can change $P_r$, consider a locked patch ahead of a rupture. Its strength exceeds the shear stress $\Sigma$. When the advancing rupture transfers an additional stress $d>0$ to the patch, the patch fails if its strength lies between $\Sigma$ and $\Sigma+d$. For a Weibull strength distribution with scale $\tau_0$ and shape $\kappa_s$, the fraction of patches that withstand a stress $s$ is $\exp[-(s/\tau_0)^{\kappa_s}]$. Dividing this fraction at $\Sigma+d$ by its value at $\Sigma$ gives the probability that a patch still locked at $\Sigma$ also withstands the stress transfer. Its probability of failure is then
\begin{equation}
	Q(\Sigma,d)=1-\exp\!\left[
	-\frac{(\Sigma+d)^{\kappa_s}-\Sigma^{\kappa_s}}
	{\tau_0^{\kappa_s}}\right].\label{eq:Q}
\end{equation}

To relate patch failure to rupture length, we assume that a barrier controls whether a rupture grows from $L$ to $r_0L$, where $r_0>1$ is a reference growth factor. If the barrier ahead of a rupture that has reached $L$ follows the strength law of a still-locked patch, the probability of reaching $r_0L$ is $P_{r_0}(\Sigma)=Q(\Sigma,d)$. With the same strength law and stress transfer at successive scales, the power-law length distribution gives
\begin{equation}
	P_r(\Sigma)=Q(\Sigma,d)^{\ln r/\ln r_0}.
\end{equation}
For $\kappa_s>1$, increasing shear stress raises the probability $Q(\Sigma,d)$ that a rupture passes a strength barrier for a given stress transfer $d$. Since $D(\Sigma)=-\ln Q(\Sigma,d)/\ln r_0$, equation~\eqref{eq:lengthb} gives
\begin{equation}
	b(\Sigma)=-\frac{3\,\ln Q(\Sigma,d)}{2\rho\,\ln r_0}.
\end{equation}
Therefore, at fixed $\rho$ and $r_0$, long ruptures become more abundant relative to short ones as shear stress increases, lowering the $b$-value.

Loading can also extend the range of rupture sizes over which the power law applies while preserving these proportions at smaller scales. In that case, $M_x$ can rise while $b$ is constant (Section~\ref{sec:regimes}).

\section{How stress accumulation and release changes the corner magnitude}\label{sec:regimes}
A stress increment can trigger a larger rupture when more neighboring patches are close to failure. For a patch that has survived stress $\Sigma$ so far, let $F_s$ be the cumulative distribution of its strength. The probability that an additional stress $d$ triggers the patch is
\begin{equation*}
 Q=\frac{F_s(\Sigma+d)-F_s(\Sigma)}{1-F_s(\Sigma)}.
\end{equation*}
The numerator counts strengths within the positive stress increment, and the denominator is the probability of surviving the initial stress. Substituting the Weibull strength distribution\cite{weibull1951} gives equation~\eqref{eq:Q}
\begin{equation}
 Q=1-\exp\!\left[-\frac{(\Sigma+d)^{k_s}-\Sigma^{k_s}}{\tau_0^{k_s}}\right].
 \label{eq:weibull}
\end{equation}
Here, $\tau_0$ is the strength scale and $k_s$ is the shape parameter of the Weibull distribution. For $k_s>1$, as expected from observations, the same additional stress triggers a surviving patch more easily as $\Sigma$ rises.

Now, let $p_c$ be the fraction of susceptible patches close to join collective failure, and define the critical stress level $\Sigma_c$ by $F_s(\Sigma_c)=p_c$. 
Expanding $p_c$ about this stress gives
\begin{equation}
 p_c-F_s(\Sigma)=F_s'(\Sigma_c)\,g+O(g^2).
\end{equation}
Here, $g=\Sigma_c-\Sigma$ represents the stress gap needed to achieve the nucleation transition.
In the critical scaling range, the characteristic number of patches in a connected cluster, $s_\xi$, and its spatial extent, $\xi$, grow as $s_\xi\propto[p_c-F_s(\Sigma)]^{-1/\sigma}$ and $\xi\propto[p_c-F_s(\Sigma)]^{-\nu}$, with positive exponents $\sigma$ and $\nu$\cite{stauffer1994,weinrib1983}. 
If the moment of a rupture involving $s$ patches scales as $s^\kappa$, its corner moment scales as $M_{0x}\propto s_\xi^\kappa$. Substituting the cluster size gives $M_{0x}\propto g^{-\kappa/\sigma}$. A change in corner moment corresponds to a change in magnitude of $\frac{2}{3}\log_{10}[M_{0x}(t)/M_{0x}(0)]$. Therefore, the corner evolves as
\begin{equation}
 M_x(t)=M_x(0)-A_x\log_{10}\!\frac{g(t)}{g_0}.
 \label{eq:regimebase}
\end{equation}
Here, $g_0$ is the initial stress gap and $A_x=\frac{2\kappa}{3\sigma}$. Differentiating with respect to time gives
\begin{equation}
 \frac{\mathrm dM_x}{\mathrm dt}=\frac{A_x}{\ln(10)}\frac{\dot\Sigma(t)}{\Sigma_c -\Sigma(t)}.
 \label{eq:cornerderivative}
\end{equation}
Therefore, the same loading rate produces faster corner growth when the remaining gap is smaller. 

Now, we consider three special cases: 
\begin{enumerate}
	\item Background seismicity: at a constant loading rate $v$, the stress gap decreases linearly as $g(t)=g_0-vt$. Substitution into equation~\eqref{eq:regimebase} gives
	\begin{equation}
		M_x(t)=M_x(0)-A_x\log_{10}(1-t/t_b).
		\label{eq:backgroundcorner}
	\end{equation}
	This expression applies from the onset of loading until the extrapolated failure time $t_b=g_0/v$, for a positive loading rate.
	\item Foreshocks: to represent feedback during mainshock preparation, let the loading rate grow with interaction length as $\dot\Sigma=v[\xi(t)/\xi(0)]^\mu$, where $v$ is the initial rate and $\mu>0$ controls the feedback. Since the interaction length scales as $g^{-\nu}$, the gap closes at rate $-\dot g=v(g_0/g)^{\mu\nu}$. Multiplying by $g^{\mu\nu}$ and integrating from the onset of loading gives
	\begin{equation*}
		\frac{g(t)^{1+\mu\nu}-g_0^{1+\mu\nu}}{1+\mu\nu}=-v g_0^{\mu\nu}t.
	\end{equation*}
	Solving for the remaining gap gives
	\begin{equation*}
		\frac{g(t)}{g_0}=(1-t/t_f)^{1/(1+\mu\nu)}.
	\end{equation*}
	Here $t_f=g_0/[(1+\mu\nu)v]$ is the time when the extrapolated gap closes. Inserting this result into equation~\eqref{eq:regimebase} gives
	\begin{equation}
		M_x(t)=M_x(0)-\frac{A_x}{1+\mu\nu}\log_{10}(1-t/t_f).
		\label{eq:regimefeedback}
	\end{equation}
	The expression applies from the onset of loading until $t_f$. Starting from the same stress and initial loading rate, feedback closes the gap sooner than steady loading.
	\item Aftershocks: following a mainshock, we let the transferred stress decay while tectonic loading continues\cite{dieterich1994}. The stress in the receiving region evolves as
	\begin{equation*}
		\Sigma(t)=\Sigma_{\rm bg}+v_at+\Delta\Sigma e^{-t/\tau}.
	\end{equation*}
	Here $t$ is time since the stress transfer, $\Sigma_{\rm bg}$ is the preceding background stress, $\Delta\Sigma$ is the transferred stress, $\tau$ is its relaxation time and $v_a$ is the non-negative reloading rate. Subtracting the stress from $\Sigma_c$ gives
	\begin{equation*}
		g(t)=g_0-v_at+\Delta\Sigma(1-e^{-t/\tau}).
	\end{equation*}
	The initial gap $g_0=\Sigma_c-\Sigma_{\rm bg}-\Delta\Sigma$ includes the mainshock stress change. Substituting the gap into equation~\eqref{eq:regimebase} gives
	\begin{equation}
		M_x(t)=M_x(0)-A_x\log_{10}\!\left[1+\frac{\Delta\Sigma(1-e^{-t/\tau})-v_at}{g_0}\right].
		\label{eq:aftercorner}
	\end{equation}
	This expression applies while the stress remains below $\Sigma_c$. For a positive stress transfer, $M_x$ falls while $(\Delta\Sigma/\tau)e^{-t/\tau}>v_a$ and rises once reloading dominates. A negative transfer lowers the initial corner and gives subsequent recovery.
\end{enumerate}
Figure~\ref{fig:s10} compares these three possible scenarios assuming a constant $b$-value.

\section{The impact of tapering and b-value changes on the probability of large earthquakes}\label{sec:probability}
Let $\lambda_0(t)$ be the rate of events above $M_c$, $\Delta M=M_t-M_c$ the magnitude difference and $R(t)=10^{1.5[M_c-M_x(t)]}(10^{1.5\Delta M}-1)$ the taper at the target $M_t$. Multiplying the event rate by the TGR exceedance probability gives $\lambda_t=\lambda_0\,10^{-b\Delta M}e^{-R}$. This relation applies to targets at or above $M_c$. Taking logarithms and using $\dot R=-1.5\ln(10)R\dot M_x$ gives
\begin{equation}
 \frac{\mathrm d\ln\lambda_t}{\mathrm dt}
 =\frac{\mathrm d\ln\lambda_0}{\mathrm dt}-\ln(10)\,\Delta M\,\dot b
 +1.5\ln(10)\,R\,\dot M_x .
 \label{eq:ratechange}
\end{equation}
At constant seismic rate, a change $\Delta b$ multiplies the GR factor by $10^{-\Delta b\,\Delta M}$, while a corner change $\Delta M_x$ replaces the initial $R$ by $R\,10^{-1.5\Delta M_x}$. The ratio of the new rate $\lambda_t'$ to the initial rate is therefore given by
\begin{equation}
 \frac{\lambda_t'}{\lambda_t}=10^{-\Delta b\,\Delta M}
 \exp\!\left[R\big(1-10^{-1.5\Delta M_x}\big)\right].
 \label{eq:multiplier}
\end{equation}
For given rates and under the Poisson hypothesis, the expected count oven target events over a horizon $T$ is $\Lambda_t(T)=\int_0^T\lambda_t(t)\,\mathrm dt$. The probability of at least one event is $P_t(T)=1-e^{-\Lambda_t(T)}$. 

Figure~\ref{fig:s7} applies this calculation to constant seismic rates. 

\paperfig{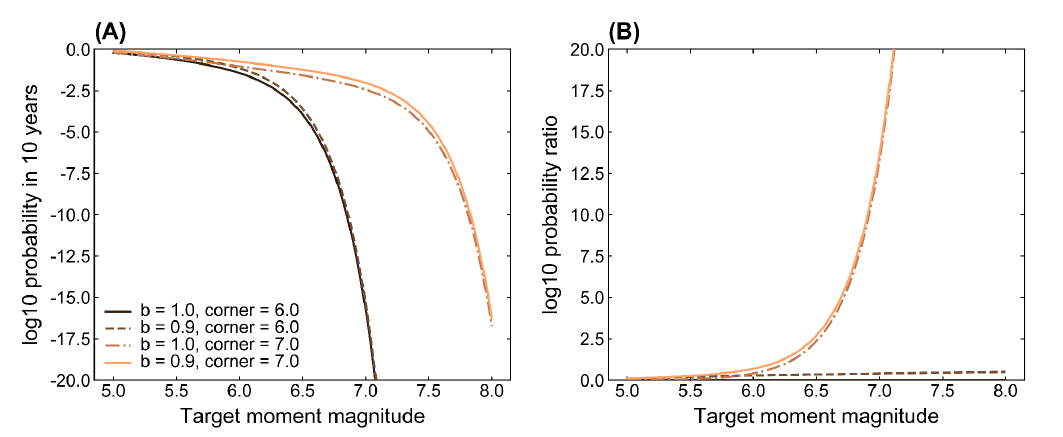}{\textbf{10-year probability increase for changes in slope and corner.} \textbf{A)} Logarithm of $P_t$ vs target moment magnitude assuming a rate of 10 events per year above $M_c=3$. \textbf{B)} Logarithm of the ratio to the probability at $b=1$ and $M_x=6$. The curves show a decrease in $b$ of 0.1, an increase in $M_x$ of one unit, and both changes together.}{fig:s7}

To isolate the effect of timing, Table~\ref{tab:regimeprob} compares a corner rising linearly from 5 to 6 over 30 days with the opposite path. With $t$ in days, we use $\lambda_0(t)=10$ events per day for background activity, $(30-t)^{-0.8}$ for accelerating foreshocks and $100/(t+0.05)$ for decaying aftershocks\cite{helmstetter2003,shcherbakov2004}. 
Reversing the corner path leaves the background probability unchanged. 
With accelerating activity, the rising corner gives a 30-day probability of $0.242\%$, compared with $0.014\%$ for the falling corner. 
With decaying activity, a high initial corner gives the larger probability ($14.6\%$ for the falling corner and $0.737\%$ for the rising corner).

\begin{table}[!htbp]\centering\small
\caption{Probability of at least one earthquake with $M\geq6$ in 30 days, for $M_c=3$ and $b=1$. The last two columns use linear corner temporal variations. Within each row the activity law is the same for all four histories.}\label{tab:regimeprob}
\setlength{\tabcolsep}{3pt}\begin{tabular}{lrrrr}\toprule Regime & $M_x=5$ & $M_x=6$ & $M_x:5\to6$ & $M_x:6\to5$\\\midrule
Background & $5.55\times10^{-15}$ & $1.04\times10^{-1}$ & $1.89\times10^{-2}$ & $1.89\times10^{-2}$ \\
Foreshocks & $1.82\times10^{-16}$ & $3.63\times10^{-3}$ & $2.42\times10^{-3}$ & $1.42\times10^{-4}$ \\
Aftershocks & $1.18\times10^{-14}$ & $2.10\times10^{-1}$ & $7.37\times10^{-3}$ & $1.46\times10^{-1}$ \\
\bottomrule\end{tabular}\end{table}\FloatBarrier

For the synthetic catalogs in Fig.~\ref{fig:s10}, let $u=t/T$ be normalised time and $s=\Sigma/\Sigma_c$ normalised stress. The three different cases reported in Section~\ref{sec:regimes} are reproduced. In all three cases we use $M_x=3.2-1.4\log_{10}[(1-s)/0.8]$ and set $b=1$. 
We sample Poisson event times with rates proportional to $1$, $(1-u+0.02)^{-0.8}$ and $(u+0.02)^{-1}$, respectively, each normalised to 2000 expected events above $M_c=2$. 
At each event time we sample a magnitude from the corresponding TGR distribution.

\paperfig{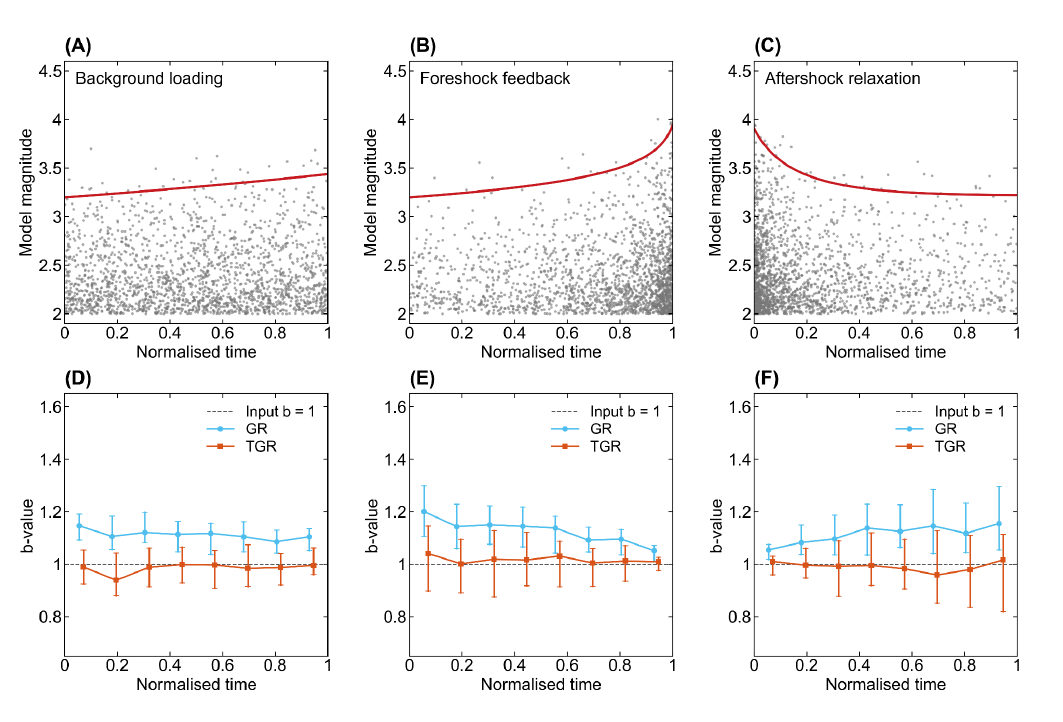}{\textbf{Magnitude distribution evolution under three different stress temporal patterns.} \textbf{A--C)} Event magnitudes (grey) and imposed corner magnitude (red) in one catalogue for background seismicity, foreshocks and aftershocks, respectively. Normalised time is the fraction $t/T$ of the simulated interval. \textbf{D--F)} GR estimates (light blue circles) and TGR slope estimates (orange squares) at different times. Points show the average values estimated over 30 simulated catalogues per each regime, and error bars represent the 1$\sigma$ uncertainties. The dashed line marks the input $b=1$.}{fig:s10}

\section{Catalog results and additional tests}\label{sec:observations}
\subsection{L'Aquila sequence and additional tests}\label{sec:laquila}
Figures~\ref{fig:s1} and~\ref{fig:s2} show the forecasts and magnitude distributions before and after the $M_w 6.3$ L'Aquila earthquake. We vary the lower magnitude threshold and the temporal weighting to test sensitivity to missed small events and to the choice of earlier earthquakes (Fig.~\ref{fig:s4}). 
Updating the slope and corner separately shows how each contributes to the forecast score (Fig.~\ref{fig:s9}).

\paperfig{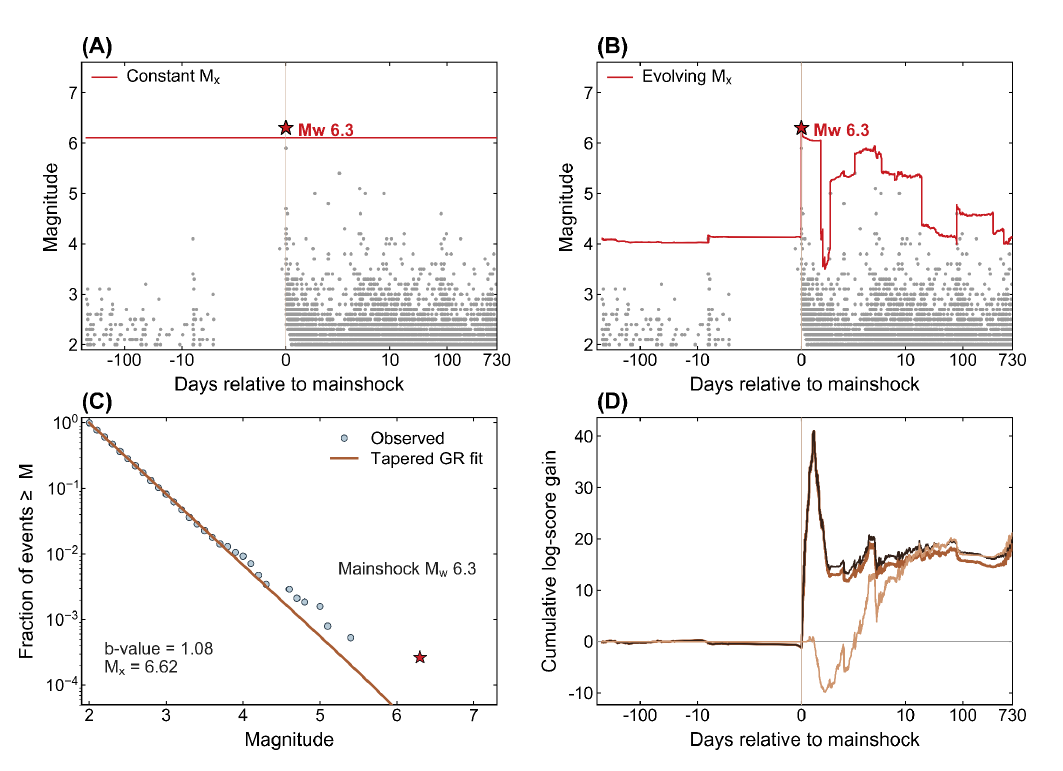}{\textbf{ETAS investigation of the evolving earthquake-size distribution at L'Aquila.} \textbf{A)} Earthquake magnitudes through the evaluation interval, with the constant corner of a TGR distribution estimated over the complete catalog shown in red. \textbf{B)} The red curve follows the corner of the TGR component estimated using up to 500 previous earthquakes. Grey points show earthquake magnitudes; the red star identifies the $M_w 6.3$ mainshock at time zero. \textbf{C)} Filled points show the cumulative fraction of earthquakes above each magnitude during the evaluation interval, including the mainshock at $M_w 6.3$. The copper curve is the TGR fit to this distribution; the red star highlights its largest event. \textbf{D)} Cumulative differences between predictive log scores between ETAS simulated catalogs generated using different best-fitting frequency-size distributions. Copper compares the updating GR-TGR model with the constant TGR reference, dark brown compares updating GR with that reference, and light copper compares GR-TGR with an expanding-history model. Results favor the first model in each comparison. Time is shown in log-scale.}{fig:s1}

\paperfig{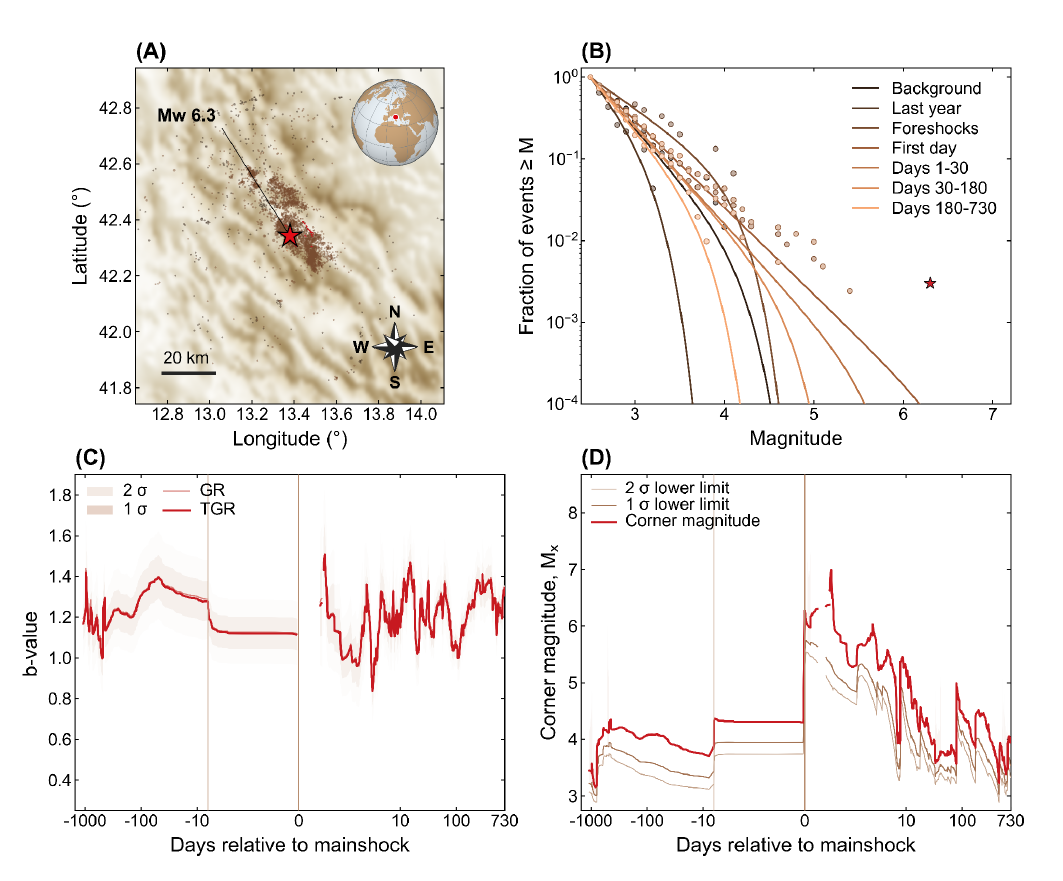}{\textbf{Seismicity in L'Aquila region.} \textbf{A)} Map of seismicity within 60 km of the mainshock. Red traces mark mapped surface faulting\cite{cinti2018database}; a red five-point star with a black boundary stands for the $M_w 6.3$ earthquake. \textbf{B)} Cumulative magnitude distributions and TGR fits for seven successive phases. Colors progress from dark brown to light copper through time; symbols are observations and curves are fits. Background covers 19 April 2005 to 6 April 2008; last year continues to the largest foreshock on 30 March 2009; foreshocks continue to the mainshock on 6 April. Recovery intervals are the first day, days 1--30, days 30--180 and day 180 to 6 April 2011. Distributions are normalised above magnitude 2.5 or the phase detection threshold. The mainshock appears at $M_w 6.3$. \textbf{C), D)} Magnitude-law parameters estimated after each event using gradual entry and exponential forgetting of past observations (see Methods). In \textbf{C)}, dark red is the TGR $b$-value and pale red the GR estimate; darker and lighter bands show nominal one- and two-standard-deviation profile support for the TGR slope ($1\sigma$ and $2\sigma$). In \textbf{D)}, red represents $M_x$ and copper curves give its lower estimates compatible with uncertainty. These temporal evolutions reveal how both the slope and the rupture scale evolve during the seismic cycle in the L'Aquila region.}{fig:s2}

\paperfig{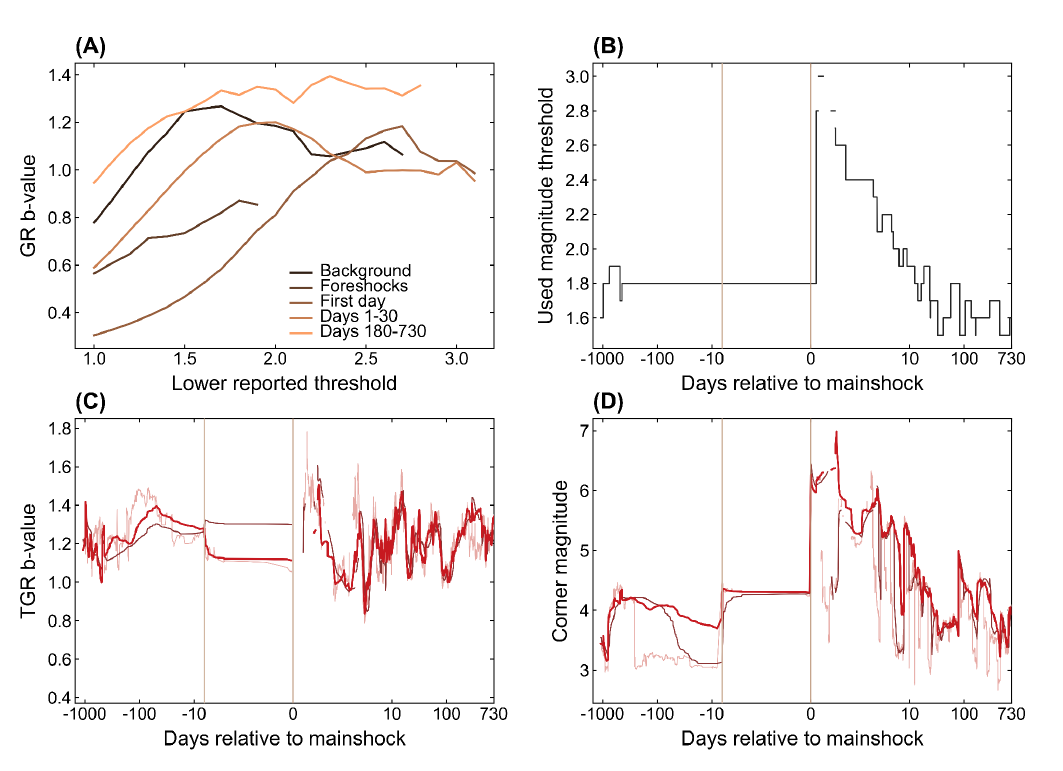}{\textbf{Threshold and temporal-weighting sensitivity at L'Aquila.} \textbf{A)} GR slopes above different magnitude thresholds for the phases in the legend. \textbf{B)} The time-dependent completeness used for the temporal evolutions. \textbf{C), D)} The corresponding slope and corner magnitude estimates from 1) rigid 300-event windows, 2) moving-average and 3) gradual event weights. Pale red, dark red and bright red identify these three different weighting schemes, respectively. Different response times explain the variation between curves, while common changes identify features that persist under the weighting choices. Time zero marks the $M_w 6.3$ mainshock.}{fig:s4}

\paperfig{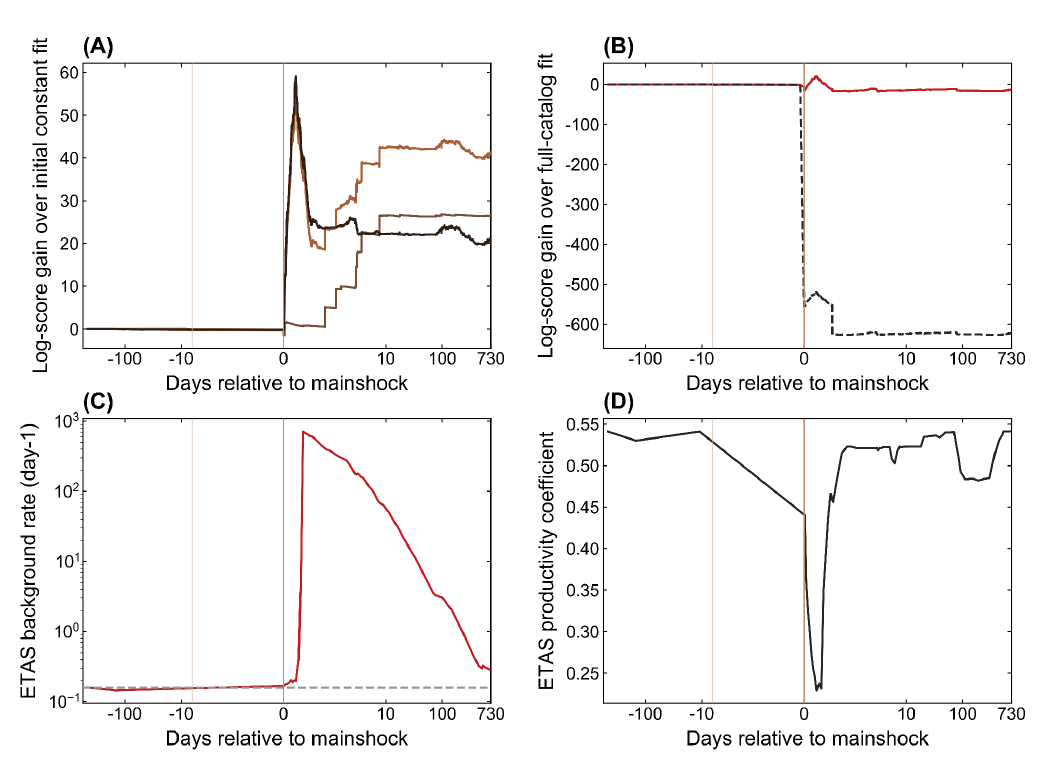}{\textbf{Information gain from different parameters to better forecast seismicity at L'Aquila} \textbf{A)} Cumulative gains for updates to both magnitude-law parameters (copper), the corner alone (brown) and the slope (b-value) alone (dark brown), relative to the initial reference distribution. \textbf{B)} Cumulative log-score gain of the evolving TGR model over a TGR model fitted once to the full L’Aquila catalogue. The dashed curve compares the two TGR models directly. For the solid red curve, each model gives 99\% weight to its own TGR distribution and 1\% to the same GR distribution. Values above zero favor the evolving model. \textbf{C), D)} ETAS background rate and ETAS productivity $K$. These rate contributions are shared by all magnitude models, so differences in \textbf{A), B)} must arise from their magnitude probabilities. Parameters are updated every 50 events.}{fig:s9}

\subsection{Ridgecrest additional analyses}\label{sec:ridgecrest}
For the $M_w7.1$ Ridgecrest sequence in Figs.~\ref{fig:forecast} and~\ref{fig:cycle}, we repeat the threshold and weighting tests (Fig.~\ref{fig:s3}) and compare forecasts that update the slope, the corner or both (Fig.~\ref{fig:s8}). Table~\ref{tab:scores} compares both catalogs using the same reference models.

\paperfig{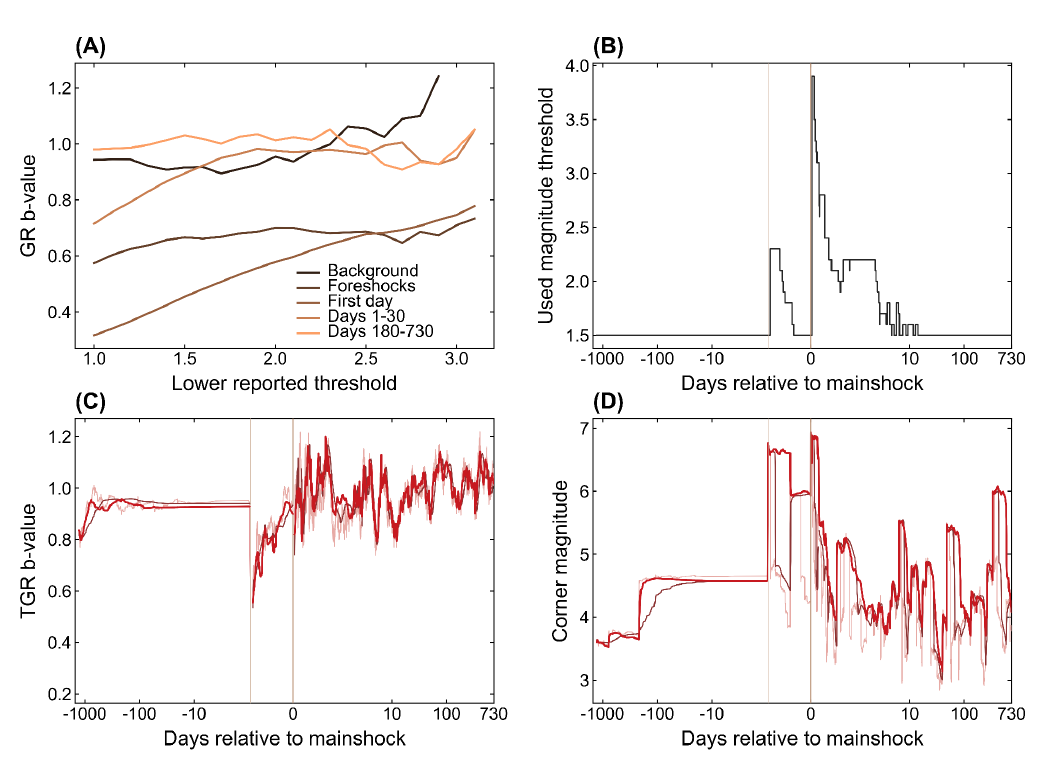}{\textbf{Threshold and temporal-weighting sensitivity at Ridgecrest.} \textbf{A)} GR $b$-values calculated above a range of lower magnitude thresholds for the selected phases named in the legend. Their variation shows the sensitivity of the slope estimate to the portion of the catalogue being fitted. \textbf{B)} The temporal evolution of $M_c$ employed for calculations. \textbf{C), D)} Slope and corner temporal evolution realized using a rigid 300-event window, a moving average and gradual event weights (see Methods). Pale red, dark red and bright red curves show the 300-event window, trailing-window average and gradual weights, respectively. The curves compare the abrupt response of a finite window with the smoother response obtained when event contributions enter and decay gradually.}{fig:s3}

\paperfig{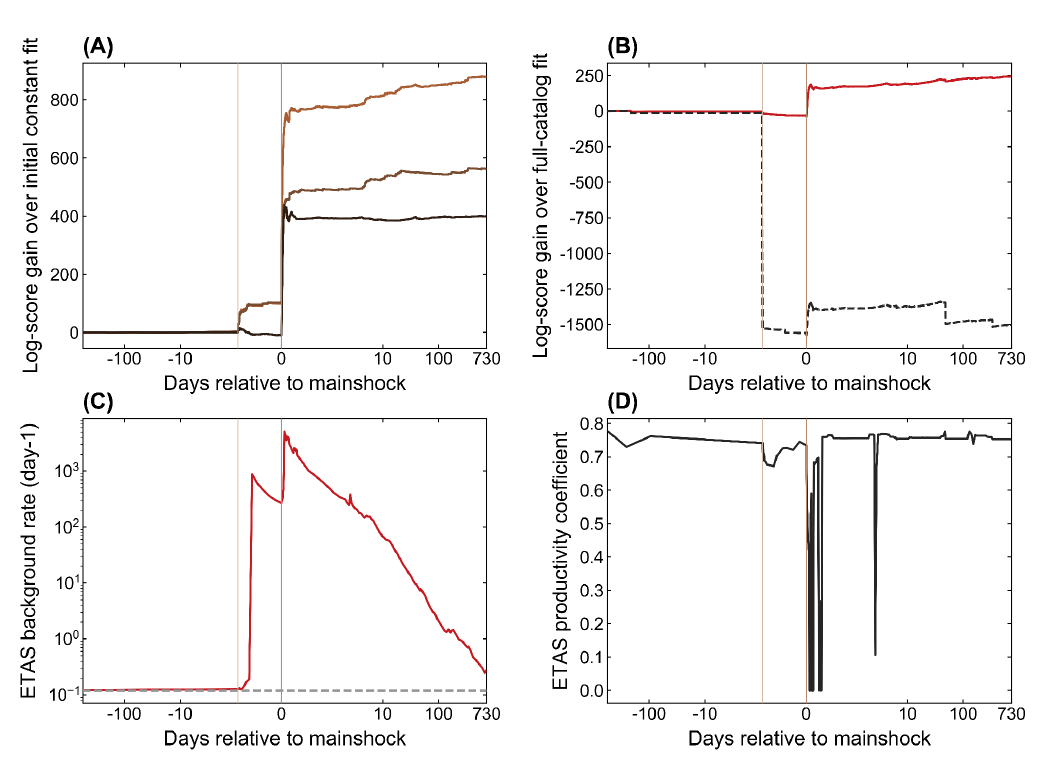}{\textbf{Information gain from different parameters to better forecast seismicity at Ridgecrest.} \textbf{A)} Cumulative gains over an initial constant magnitude law when both $b$ and $M_x$ (copper), only $M_x$ (dark brown), or only $b$ (brown) updated every 50 events. Each predictive distribution includes the possibility of degeneracy from TGR to GR. \textbf{B)} Updating both parameters compared with the full-catalog constant TGR reference. The red curve shows the 1\% mixture and the dashed dark curve shows pure TGR. \textbf{C), D)} ETAS background rate and productivity $K$ through the same interval. These rate contributions are shared by all magnitude models, so differences in \textbf{A), B)} must arise from their magnitude probabilities. Time zero marks the $M_w 7.1$ mainshock.}{fig:s8}

\begin{table}[!htbp]\centering\small
\caption{ETGR model score gains against three references. Positive gains favor the time-evolving model. Initial and expanding references use all the previous events. The full-catalog reference is a constant TGR fit.}\label{tab:scores}
\begin{tabular}{llrrr}\toprule Sequence & Interval & Initial & Expanding & Full catalogue\\\midrule
L'Aquila & Complete interval & $3.36\times10^{1}$ & $2.12\times10^{1}$ & $1.88\times10^{1}$ \\
L'Aquila & Before mainshock & $1.50\times10^{-1}$ & $0$ & $-7.34\times10^{-1}$ \\
Ridgecrest & Complete interval & $5.79\times10^{2}$ & $3.38\times10^{2}$ & $3.31\times10^{2}$ \\
Ridgecrest & Before mainshock & $5.04\times10^{1}$ & $-5.31\times10^{0}$ & $-9.34\times10^{0}$ \\\bottomrule\end{tabular}\end{table}\FloatBarrier

\section*{GR bias calculation}
\paperfig{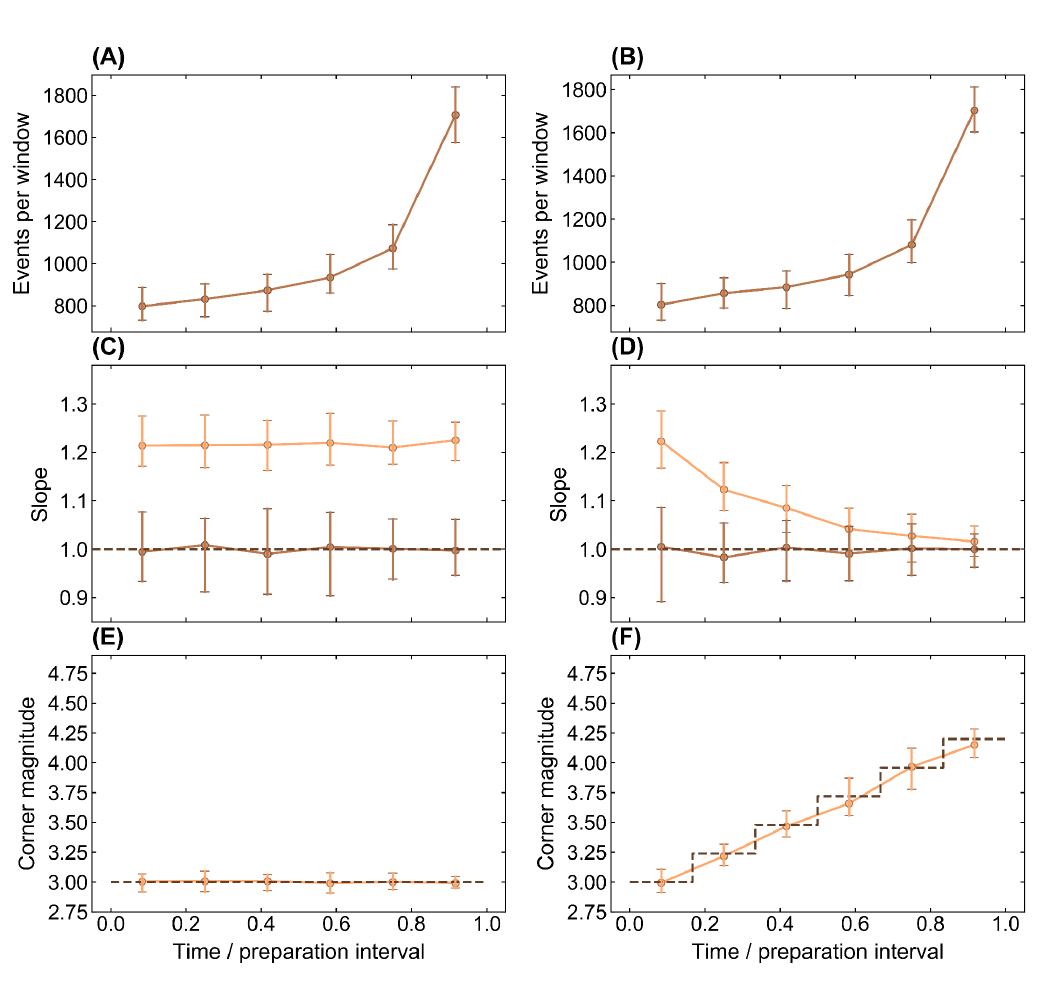}{\textbf{Estimation of the b-value in a simulated TGR ETAS catalog with constant (left) and changing (right) corner magnitude}. The true $b=1$. The input corner magnitude is fixed at $M_x =$ 3.0 (left) or increases from 3 to 4.2 (right). \textbf{A, B)} Event counts through time. \textbf{C, D)} GR slopes (light copper), fitted TGR slopes (medium copper) and input (true) slope (dashed dark brown). \textbf{E, F)} Fitted (light copper) and input (dashed dark brown) corner magnitudes. Error-bars represent 2$\sigma$ uncertainties.}{fig:s5}
\end{document}